\documentclass[reprint,amsmath,amssymb,english,aps,pra,superscriptaddress]{revtex4-1}

\usepackage[T1]{fontenc}
\usepackage{amsmath}
\usepackage{amssymb}
\usepackage{graphicx}
\usepackage{babel}
\usepackage{color}
\usepackage{braket}
\usepackage[normalem]{ulem}
\usepackage[colorlinks,bookmarks=false,citecolor=magenta,linkcolor=blue,urlcolor=black]{hyperref}
\usepackage{xcolor}

\usepackage{amssymb}
\usepackage{amsmath}
\usepackage{amsthm}
\usepackage{amsfonts}
\usepackage{braket}
\usepackage[normalem]{ulem}
\usepackage{lineno}

\begin{document}


\title{Response of macroscopic and microscopic dynamical quantifiers to the quantum critical region}

\author{Stav Haldar}
\affiliation{Harish-Chandra Research Institute, HBNI, Chhatnag Road, Jhunsi, Allahabad 211 019, India}
\affiliation{Hearne Institute for Theoretical Physics and Department of Physics and Astronomy, Louisiana State University, Baton Rouge, LA USA}

\author{Saptarshi Roy}
\affiliation{Harish-Chandra Research Institute, HBNI, Chhatnag Road, Jhunsi, Allahabad 211 019, India}

\author{Titas Chanda}
\affiliation{Instytut Fizyki Teoretycznej, Uniwersytet Jagiello\'nski,  \L{}ojasiewicza 11, 30-348 Krak\'ow, Poland}

\author{Aditi Sen(De)}
\affiliation{Harish-Chandra Research Institute, HBNI, Chhatnag Road, Jhunsi, Allahabad 211 019, India}


\begin{abstract}
At finite temperatures, the quantum critical region (QCR) emerges as a consequence of the interplay between thermal and quantum fluctuations. 
We seek for suitable physical quantities, which during dynamics can give prominent response to QCR in the transverse field quantum $XY$ model.
We report that the maximum energy absorbed, the nearest neighbor entanglement and the quantum mutual information of the time evolved state after a quench of the transverse magnetic field exhibits a faster fall off with temperature when the initial magnetic field is taken from within the QCR, compared to the choice of the initial point from different phases.
We propose a class of dynamical quantifiers, originated from the response of these physical quantities and show that they can faithfully mimic the equilibrium physics, namely detection of the QCR  at finite temperatures.



\end{abstract}

\maketitle

\section{Introduction}

The onset of a phase transition is signalled by the emergence of long-range fluctuations in the system. In the classical domain, these fluctuations are solely driven by temperature leading to a phase transition when the temperature crosses a critical value.  On the other hand, 
in the absolute zero temperature, where thermal fluctuations are completely absent, quantum phase transitions (QPTs) \cite{qptbook4, qptbook2, qptbook3, qptbook1} can occur, which are exclusively driven by quantum fluctuations.
Such transition typically takes place at specific value(s) of system parameter(s) called the quantum critical point(s) (QCP). 
The QCP in a QPT takes the analogous role of the critical temperature in case of classical phase transitions.
It was recently found that the presence of QCP can interestingly induce nonanalyticities in the dynamics of physical quantities after a quench of system parameters in paradigmatic one-dimensional quantum spin models. The study of these dynamical nonanalyticities goes by the name of dynamical quantum phase transition (DQPT) \cite{KS2004, ASD2005, pollmann, heyl_prl, heyl_prl_2, heyl_prl_3, heylprl4, heyl_review,  heyl_cont_sym_break, Vajna_prb}, which was shown to be detected by Loschmidt echo \cite{heyl_review} and entanglement measures \cite{schmidtgap,dqpt-stav1}.

Typically, studies of phase transition hovers around two extremes: the high temperature (classical) limit and the zero temperature quantum limit. However, the story becomes interesting when one probes in regimes where both thermal and quantum fluctuations are finite, and comparable. In such a domain, the conventional knowledge of physics near the critical points is not of much help since that region does have a different phase. Therefore,  a complete understanding of a system where both of these fluctuations co-exist is an essential challenge that is yet to be resolved in full generality. 
Nevertheless, there has been a number of efforts to uncover the region, known as quantum critical region (QCR) \cite{qptbook2, qptbook1}, where both thermal and quantum effects are present. In the finite temperature domain, studying the canonical equilibrium state reveals that the quantum critical point expands into a conical region, the QCR, with the zero temperature QCP at its vertex. 
Note that the QCR does not correspond to a well defined phase, bounded not by critical lines, but rather by crossover regions. Therefore,
to detect such a region, no order parameter based classification or gap closing criterion exists, thereby making its identification extremely challenging. 
Nevertheless, several investigations are carried out to characetrize  QCR \cite{qd1, qd2, qd3, qd4, qd5, qd6, qd7, qd8}. Furthermore, studies of the same using the information-theoretic quantities include detection through directional derivatives of magnetization and entanglement \cite{Amico2008},  Benford analysis of magnetization \cite{AmeyaPRE}. Recently, the coveted QCR was also  observed experimentally \cite{sachdevPRX}.
Note that the properties of the QCR are dictated by scale invariance which it borrows from the underlying QCP. The boundaries of the QCR for low temperatures is defined by the straight lines 
\begin{eqnarray}
T \approx C \vert h-h_c \vert,
\label{qcr}
\end{eqnarray}  
where $C$ is a constant determined by the universality class of the system \cite{qptbook2}, and $h$ represents the system parameter, while $h=h_c$ denotes the QCP.

 \begin{figure}[t]
 \includegraphics[width=0.7\linewidth]{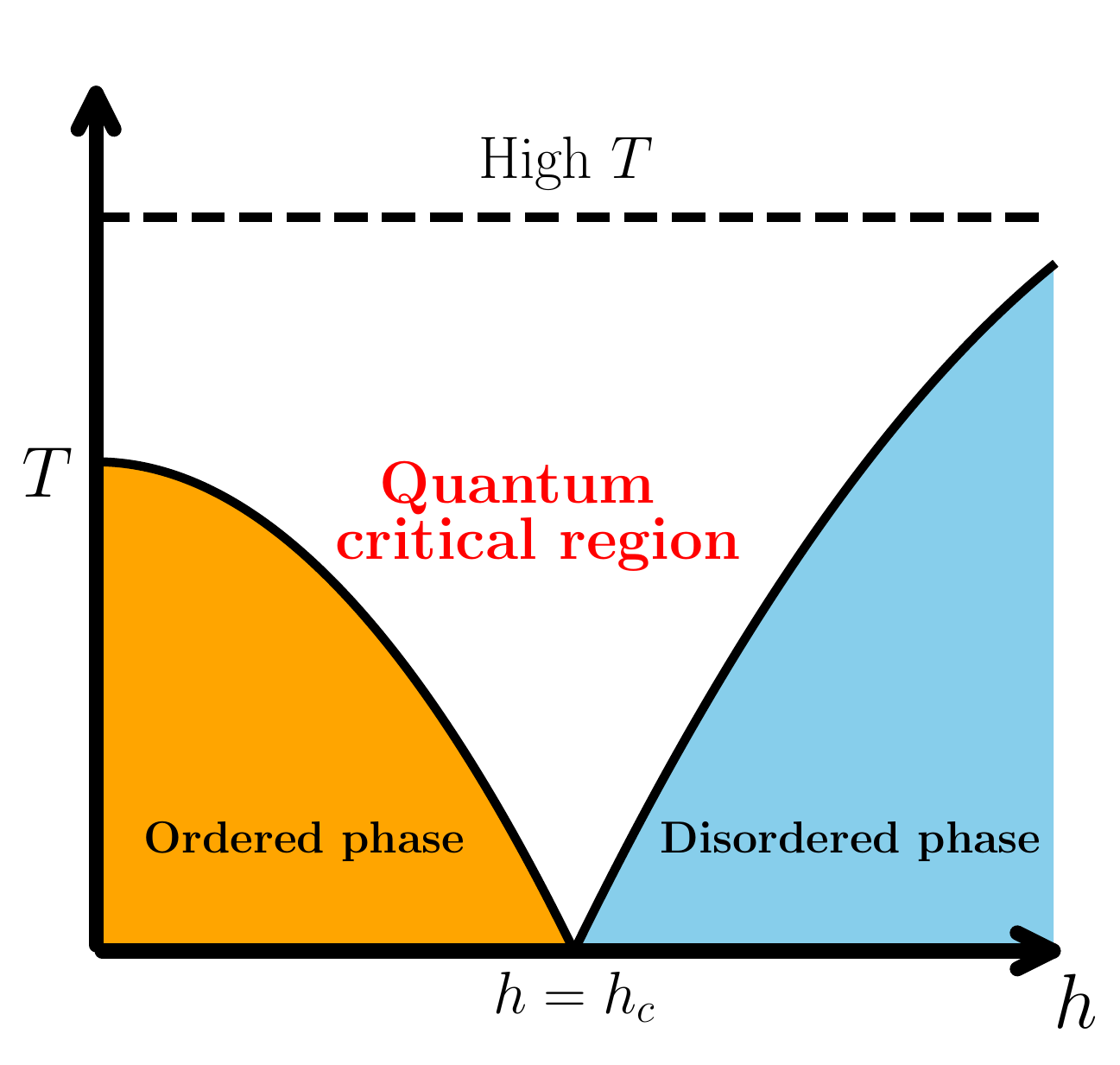}
 \caption{(Color online.) A schematic representation of the quantum critical region (QCR), the ordered and disordered phases with respect to the system parameter $h$ and temperature $T$. At zero temperature, the system undergoes a quantum phase transition from ordered to disordered phase at $h=h_c$. At high temperatures, strong thermal fluctuations typically washes out the quantum phases.}
 \label{fig:qcr-schematic}
 \end{figure}

   In this paper, we search for physical characteristics at finite temperatures
   which yield a well defined response to the QCR during their \emph{out of equilibrium dynamics}.  We believe, investigations of  QCR during evolution set our analysis different from most of the other studies which typically  deal with the equilibrium setting. In particular,
 we report that energy absorbed during a time pulse, a quantity of macroscopic origin, and microscopic information theoretic quantities like bipartite entanglement and quantum mutual information, manifest a qualitative difference in their dynamics when the initial quench point is ``within'' the QCR. Interestingly, the dynamical response of these quantities can be structured in a way which enables quantitative demarcation of the QCR.  Notice that 
 since there exists an intrinsic fuzziness in the boundaries of QCR even in equilibrium, our analysis also observes such  inherent uncertainity during the detection of QCR from the dynamics of the physical quantities.
 
The paper is organized as follows: Sec. \ref{sec:model} describes the system we consider for our analysis and its critical properties. In Sec. \ref{sec:signature}, we show that temperature dependence of macroscopic (Sec. \ref{subsec:macro}) and microscopic (Sec. \ref{subsec:micro}) dynamical quantifier can  indicate  QCRs during out of equilibrium dynamics. Based on the analysis of Sec. \ref{sec:signature}, we quantitatively demarcate QCRs in Sec. \ref{sec:marking} from ordered and disordered regions. Finally, we draw conclusions in Sec. \ref{sec:concl}.

\section{The model and its criticalities}
\label{sec:model}

We consider an interacting quantum spin-$1/2$ system on a one-dimensional (1D) lattice with nearest-neighbor anisotropic $XY$ interaction in presence of a uniform external transverse magnetic field. The model is described by the Hamiltonian 
\begin{eqnarray}
\hat{H} &=& 
\frac{1}{2}\sum_{j=1}^N \Big[J \big(\frac{1+\gamma}{2}\hat{\sigma}^x_j\hat{\sigma}^x_{j+1} + \frac{1-\gamma}{2}\hat{\sigma}^y_j\hat{\sigma}^y_{j+1}\big)
+h \hat{\sigma}^z_j\Big] \nonumber \\
\label{eq:hamil}
\end{eqnarray}
with periodic boundary condition.
Here, $\hat{\sigma}^\alpha$  with $\alpha = x, y, z$ are the Pauli matrices, $J$ represents the strength of nearest-neighbor exchange interaction,  $\gamma$ is the anisotropy parameter in the $x-y$ direction, $h$ is the uniform transverse magnetic field strength, and $N$ denotes the total number of lattice sites. 
The above Hamiltonian can be mapped to a non-interacting 1D spinless Fermi system via the Jordan-Wigner transformation, and then can be solved exactly via successive Fourier and Bogoluibov transformations \cite{bm1} (see Appendix \ref{Appendix} for details). For non-zero choices of $\gamma$, the 1D quantum $XY$ model at zero temperature has a QCP, belonging to the Ising universality class, at $h/J = \pm 1$ 
where the model undergoes from an ordered phase, a ferromagnetic one for $J<0$ or an antiferromagnetic one for $J>0$ (with $|h|<|J|$) to a disordered paramagnetic phase (when 
$|h| >|J|$) with critical exponents $z=1$, $\nu=1$ \cite{qptbook1}.

As mentioned earlier, for non-zero temperatures, the QCPs expand into a conical shaped region, called the QCR, see Fig. \ref{fig:qcr-schematic}. 
Here, we set out to investigate how the dynamics gets effected in a finite temperature situation due to the presence of QCR via both macroscopic and microscopic quantities. In our analysis for the Ising transition, we fix $\gamma=0.8$ unless otherwise stated. The results remain qualitatively similar for different choices of $\gamma \neq 0$ as well.

We know that apart from the Ising transition, the above model undergoes an anistropic phase transition, where the (anti)ferromagnetic ground state changes its orientation from $x$ to $y$ direction or vice versa, across the critical line $
(\gamma = 0, |h| < |J|)$ having the same critical exponents as the Ising transition. The model also displays a multicritical transition at the point where the Ising and anisotropic critical lines meet, i.e., at $(\gamma = 0, h =\pm J)$, which has different critical exponents ($z=2$ and $\nu=1/2$) than the former ones \cite{qptbook1}. We also repeat our analysis for this multicritical point, where we consider $\gamma = 1 - |h(t=0)/J|$.

\section{Signatures of Quantum Critical Region: Out-of-equilibrium Dynamics}
\label{sec:signature}

In this section, we study the response of physical quantities both macroscopic (absorbed energy) and microscopic information theoretic quantities (bipartite entanglement and mutual information) after a quench of the magnetic fields in the presence or absence of the quantum critical region.

\subsection{Macroscopic signatures of QCR: Falloff behavior of energy}
\label{subsec:macro}

In the context of zero temperature QPT, it was shown \cite{Subinoy} that the energy absorbed in a square pulse quench of the magnetic field (see below), in the long time limit develops some kinks when the quench crosses an equilibrium quantum critical point of the transverse field $XY$ model. Therefore, this intrinsic non-equilibrium quantity could mimic equilibrium properties. 
This motivated us to investigate the features of this quantity at finite temperatures, and search for signatures of the QCR.
We study dynamics of the canonical equilibrium state of temperature $T$, under a time pulse of the external transverse magnetic field of the form,
\begin{eqnarray}
  h(t)= \left\{
 \begin{array}{cc}
 h_0, & t\leq 0,  \\
 h_1, & \ \ \ \ \ 0 < t \leq \tau, \\
 h_0, & t > \tau,
\end{array}\right.
     	 \label{eq:quench}
\end{eqnarray}
i.e., the initial Hamiltonian $\hat{H}_0$ corresponding to a transverse field strength $h_0$ is quenched to a new Hamiltonian $\hat{H}_1$ with transverse field strength $h_1$. The time evolution of the thermal state of $\hat{H}_0$ at a temperature $T$ is allowed to take place with $\hat{H}_1$ for a finite time $\tau$. Finally, the external field is quenched back to its original value $h_0$. 
We follow the same quench strategy for analyzing  both the Ising criticality and multicritical transition.
Since the $XY$ model is integrable, the time evolved state at time $\tau$ can be calculated analytically, and allows one to evaluate the energy absorbed during the time pulse, which is defined as
\begin{eqnarray}
\Delta E(T, \tau) = \text{Tr}(\hat{H}_0 e^{-i \hat{H}_1 \tau} \hat{\rho}_0 e^{i \hat{H}_1 \tau}) - \text{Tr}(\hat{H}_0 \hat{\rho}_0),
\label{eq:e1}
\end{eqnarray} 
where $\hat{\rho}_0 = e^{-\frac{\hat{H}_0}{k_B T}}/\text{Tr }e^{-\frac{\hat{H}_0}{k_B T}}$, with $k_B$ being the Boltzmann constant, represents the thermal state. 

\begin{figure*}[t]
    \centering
    \includegraphics[width=0.9\linewidth]{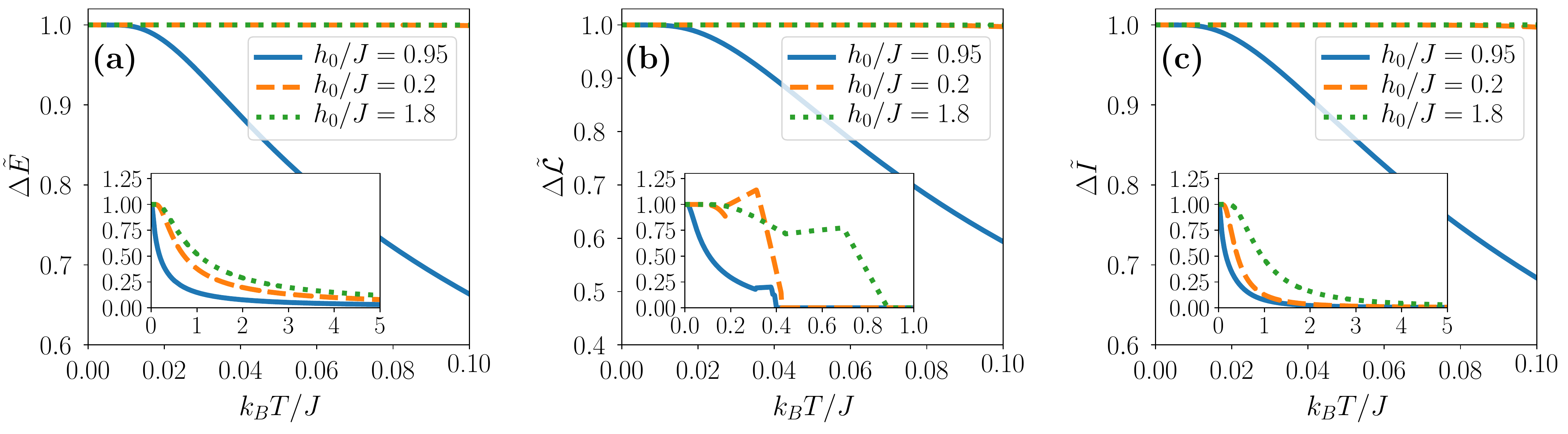}
    \caption{(Color online.) Temperature dependence of $\Delta \tilde{\mathcal{Q}} = \frac{\Delta \mathcal{Q}_{\max}(T)}{\Delta \mathcal{Q}_{\max}(T=0)}$ for a given physical quantity, $\mathcal{Q}$. The abscissa represent $k_BT/J$ and the ordinate is $\Delta\tilde{\mathcal{Q}}$. (a) $\mathcal{Q}$ is the energy absorbed during a time pulse, (b)  $\mathcal{Q}=$ logarithmic negativity, (c) $\mathcal{Q}=$ quantum mutual information. Plots corresponds to the analysis of Ising criticality, i.e., transverse field $XY$ model with $\gamma=0.8$, and
 are for different values of $h_0/J$ and a fixed value of the final quench parameter $h_1/J=1$.  For $k_BT/J \lesssim 0.1$, 
         $\Delta \tilde{\mathcal{Q}}$ falls sharply when the quench begins from the QCR whereas it is almost independent of temperature when the quench begins from the ordered and disordered regimes. For higher temperatures, $\Delta \tilde{E}$ and $\Delta \tilde{\mathcal{I}}$ reach zero asymptotically, as depicted in inset (a) and (c), while $\Delta \tilde{\mathcal{L}}$ displays nonmonotonicity and sudden collapse with respect to $T$, shown in inset (b). All axes are dimensionless.}
    \label{fig:E_abs}
\end{figure*}

\begin{figure}
\includegraphics[width=0.65\linewidth]{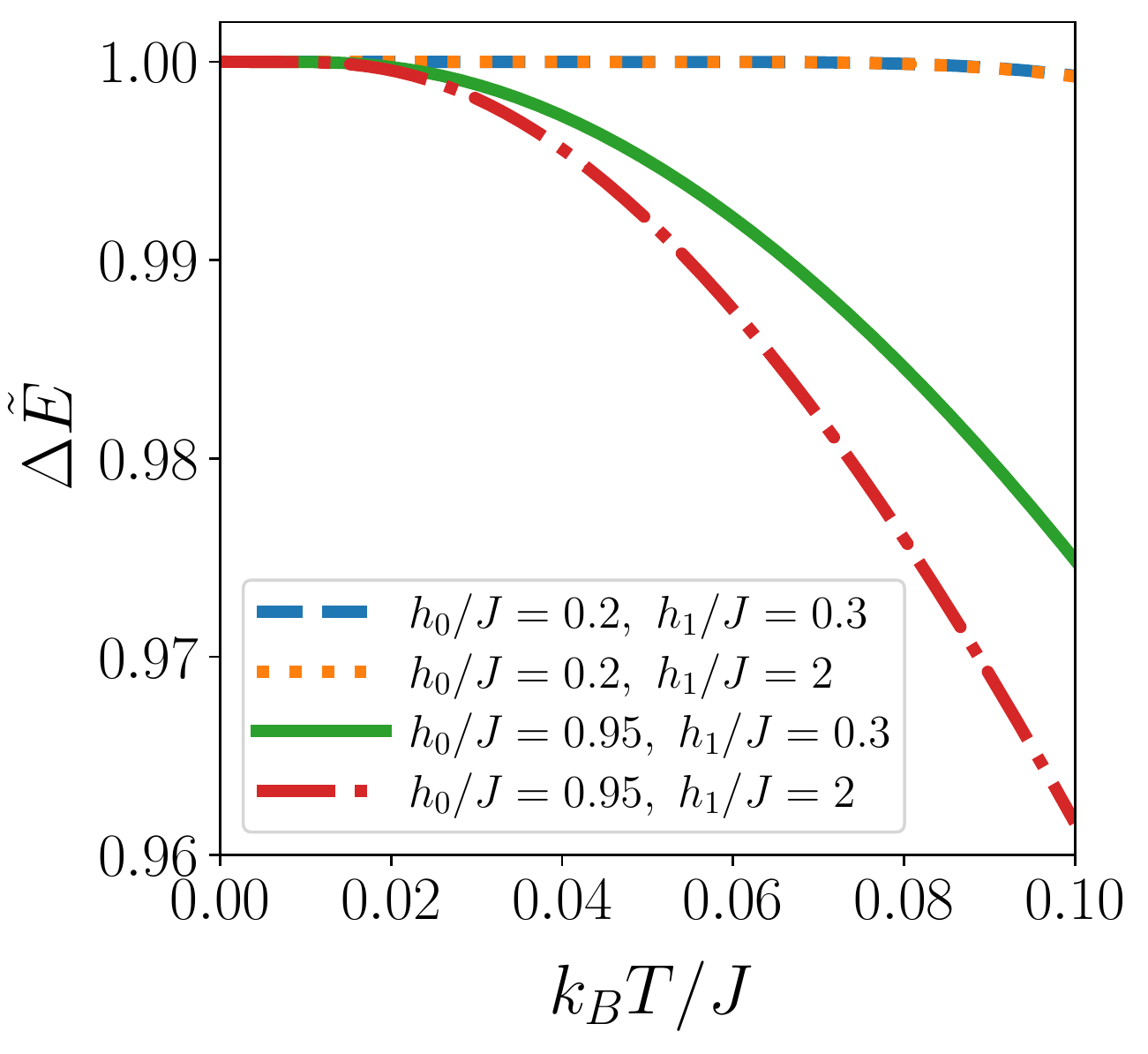}
\caption{(Color online.) Independence of the fall off feature in $\Delta \tilde{E}$ for quantum $XY$ model with transverse field and $\gamma=0.8$.
The abscissa represents temperature, $k_BT/J \lesssim 0.1$, while the ordinate represents $\Delta \tilde{E}$. 
$\Delta \tilde{E}$ is calculated for different choices of $(h_0/J,h_1/J)$-pair.
 The behavior of $\Delta \tilde{E}$ clearly shows that the characteristic temperature dependence is independent of the quench length $\vert h_0/J - h_1/J \vert$. Both axes are dimensionless.}
\label{h0_scaling}
\end{figure}  

The value of the absorbed energy maximized over the duration of the pulse can be calculated as
\begin{eqnarray}
\Delta E_{\max}(T) = \max_\tau \vert \Delta E(T, \tau) \vert.
\end{eqnarray}
In order to reveal the equilibrium QCR, we analyze the temperature dependence of the scaled quantity
\begin{eqnarray}
\Delta \tilde{E}(T) = \frac{\Delta E_{\max}(T)}{\Delta E_{\max}(T=0)}.
\end{eqnarray} 
The above quantity depends both on the initial and final fields ($h_0$ and $h_1$) and consequently on their corresponding quantum  phases.
To obtain proper response for equilibrium QCR,
the final quench point should be chosen suitably such that it maximizes the distinct temperature dependence of $\Delta \tilde{E}(T)$. We find that choosing the final quench point as $h_1/J = 1$ gives rise to strong temperature dependence of $\Delta \tilde{E}(T)$ for almost any choice of the initial quench point $h_0$, and hence we fix $h_1/J = 1$ for all our quantitative analyses. However, the qualitative features reported here remains unaltered even for different values of the final field strength  $h_1$ as we shall see later.
Therefore, we are left with a quantity that effectively depends on only the $(h_0,T)$-pair, which will be used to demarcate the QCR. 

For now we restrict ourselves only to the analysis of Ising criticality, and the qualitative findings (see Fig. \ref{fig:E_abs} (a)) of our investigation corresponding to the Ising transition using $\Delta \tilde{E}(T)$, for $k_B T/J \lesssim 0.1$, is summarized below (note that the behavior is qualitatively same for multicritical transition):
\begin{enumerate}
\item If the initial field strength, $h_0$, is chosen from deep inside the ordered or disordered phases, $\Delta \tilde{E}(T)$ is an almost constant function of $T$, i.e., it changes negligibly with changing temperature.
\label{pt1}

\item If $h_0$ is inside the quantum critical region, i.e. from points close to the quantum critical point, there is a rapid fall of $\Delta \tilde{E}(T)$ with temperature. The closer the initial quench point is to the QCP, the faster the fall.  
\label{pt2}

\item The relevant physics reported in \ref{pt1} and \ref{pt2} above is independent of the quench length, $|h_0/J - h_1/J|$, in the sense that they do not effect the fall off feature in $\Delta \tilde{E}(T)$. For example, for the quench: $ \frac{h_0}{J} (0.2) \rightarrow \frac{h_1}{J} (0.3, 2)$, we get that all the relevant quantities under consideration remain almost constant for $\frac{k_BT}{J} \lesssim 0.1$, while for the quench: $\frac{h_0}{J} (0.95) \rightarrow \frac{h_1}{J} (0.3,2)$, we observe a rapid fall off in $\Delta \tilde{\mathcal{Q}}$.
See Fig.   \ref{h0_scaling}. 
As mentioned earlier, the choice of $h_1=J$ was motivated by the observation that driving by the Hamiltonian with critical parameters yield a strong response to the QCR. Similar behavior can also be seen with microscopic quantities,  which are discussed later in the manuscript.
\label{pt3}

\end{enumerate}
On the other hand, for higher temperatures, the thermal fluctuations surpass its quantum counterparts and the signatures of QCR vanish beyond a certain temperature. In particular, for any quenches beginning either in ordered, disordered or in QCR, the physical quantity of interest decreases in a similar manner with increasing temperature and vanishes asymptotically, see inset of Fig. \ref{fig:E_abs} (a).

\subsection{Microscopic signatures of QCR: Entanglement and mutual information}
\label{subsec:micro}

Let us now consider the dynamics of two microscopic quantities under the quenching strategy,
\begin{eqnarray}
h(t)= \left\{
 \begin{array}{cc}
 h_0, & t\leq 0  \\
 h_1, & t>0
\end{array}\right..
\label{eq:quench2}
\end{eqnarray}

We start of our analysis with bipartite entanglement \cite{hhhh}, which have been known to play an important role in detecting equilibrium QCPs \cite{Fazioqpt, Osborne2002, rmp-amico, aditi-mac, mac-book, atxy_pra, amader_dm}, although it turns out to be less effective in the case of dynamical phase transition at zero temperature \cite{dqpt-stav1}.
However, in thermal equilibrium, entanglement has been already used  to successfully detect the QCR \cite{Amico2008}. Therefore, it is important to explore  dynamical response of entanglement due to QCR.

To quantify entanglement, we use negativity ($\mathcal{N}$) and logarithmic negativity ($\mathcal{L}$) \cite{Peres, negativity, ln}, which are defined for an arbitrary two party density matrix $\rho_{AB}$,  as
\begin{eqnarray}
\mathcal{N}(\hat{\rho}_{AB}) &=& \frac{1}{2}(||\hat{\rho}^{T_B}_{AB}|| - 1) = \frac{1}{2}(||\hat{\rho}^{T_A}_{AB}|| - 1), \nonumber \\ 
\mathcal{L}(\hat{\rho}_{AB}) &=& \log_2(2\mathcal{N}(\hat{\rho}_{AB})+1),
\end{eqnarray}
where $||A|| = \text{Tr}\sqrt{A^\dagger A}$ and $T_{A}(T_{B})$ denotes partial transposition with respect to party $A(B)$ \cite{ap-sep,necess-suff}. 
The dynamics of nearest-neighbor negativity as well as logarithmic negativity after a sudden quench as in Eq. \eqref{eq:quench2}, can be computed analytically
in terms of classical correlators $C^{ij}=\text{ Tr}\big(\hat{\rho}_{AB}(t,T)\sigma^i \otimes \sigma^j\big)$ with $i,j=x,y,z$,  and the magnetization in the $z$ direction, $m^z  =\text{ Tr}\big(\hat{\rho}_{AB}(t,T) \mathbb{I}_2 \otimes \sigma^z \big)=\text{ Tr}\big(\hat{\rho}_{AB}(t,T)\sigma^z \otimes \mathbb{I}_2\big)$, 
where $\hat{\rho}_{AB}(t, T)$ denotes the density matrix corresponding to nearest-neighbor sites at a given time $t$ and temperature $T$ 
(see appendix \ref{Appendix} for details). 
The exact diagonalization of $\hat{H}$ guarantees the analytical forms of the nonvanishing classical correlators and magnetization.
Note that owing to the translational invariance of the model, all nearest-neighbor density matrices of the initial as well as the time evolved states, $\hat{\rho}_{AB}(t,T)$, are identical.

Like in the case of energy absorbed, we consider the maximal value of change in logarithmic negativity, which is defined as
\begin{eqnarray}
\Delta\mathcal{L}_{\max}(T) = \max_t \big\vert \mathcal{L}(T,t)-\mathcal{L}(T,t=0)\big\vert,
\end{eqnarray}
and  the corresponding scaled quantity as
\begin{eqnarray}
\Delta \tilde{\mathcal{L}}(T)=\frac{\Delta\mathcal{L}_{\max}(T)}{\Delta\mathcal{L}_{\max}(T=0)}.
\end{eqnarray}
For low temperatures, $k_BT/J \lesssim 0.1$, $\Delta \tilde{\mathcal{L}}(T)$ also remain constant when the initial field strength is taken far from QCP while it decreases with $k_BT/J$ when $h_0/J$ is chosen from the QCR (as depicted in Fig. \ref{fig:E_abs} (b) for Ising criticality). 
However, unlike $\Delta \tilde{E}(T)$,  we notice that above $k_BT/J > 0.1$,  $\Delta \tilde{\mathcal{L}}(T)$ reveals non-monotonic behavior, see Fig. \ref{fig:E_abs} (b), as reported in various earlier works in the static scenario \cite{atxy_pra, amader_dm}. Furthermore, at high temperatures, $\Delta \tilde{\mathcal{L}}(T)$ suddenly collapses to vanishingly small values which is not the case for $\Delta \tilde{E}(T)$, which rather approaches zero asymptotically (compare insets of Figs. \ref{fig:E_abs} (a) and \ref{fig:E_abs} (b)).

 
We move on with our search for physical quantities and consider quantum mutual information \cite{CoverThomas} for the same. Unlike entanglement, which solely measures the quantum correlations, mutual information, $\mathcal{I}_{A:B}$, is a measure of the total correlations between the parties $A$ and $B$ \cite{totalcorr1,totalcorr2}.  For a bipartite density matrix $\hat{\rho}_{AB}$, the mutual information  is defined as
\begin{eqnarray}
\mathcal{I}_{A:B} = S(\hat{\rho}_{A}) + S(\hat{\rho}_{B}) - S(\hat{\rho}_{AB})
\label{eq:mi-def}
\end{eqnarray}
where $S(\hat{\sigma}) = -\text{Tr } [\hat{\sigma} \log_2 \hat{\sigma}]$ is the von Neumann entropy of a state $\hat{\sigma}$, and  $\hat{\rho}_{A(B)}$ is the reduced density matrices of $\hat{\rho}_{AB}$.   
Like in the previous cases, the mutual information of $\hat{\rho}_{AB}(t,T)$ can be evaluated  analytically for the system considered (see appendix \ref{Appendix} for details).
The maximal change in mutual information and its scaled variant for a given temperature $T$  are defined respectively as
\begin{eqnarray}
& &\Delta \mathcal{I}_{\max}(T) = \max_t  \big\vert \mathcal{I}_{A:B}(T,t) - \mathcal{I}_{A:B}(T,t =0) \big\vert, \nonumber \\
& &\Delta \tilde{\mathcal{I}}(T) = \frac{\Delta\mathcal{I}_{\max}(T)}{\Delta\mathcal{I}_{\max}(T=0)},
\end{eqnarray}
which again decreases with the variation of temperature when $h_0 \in$ QCR, as depicted in Fig. \ref{fig:E_abs} (c).



\begin{figure}[h]
\includegraphics[width=\linewidth]{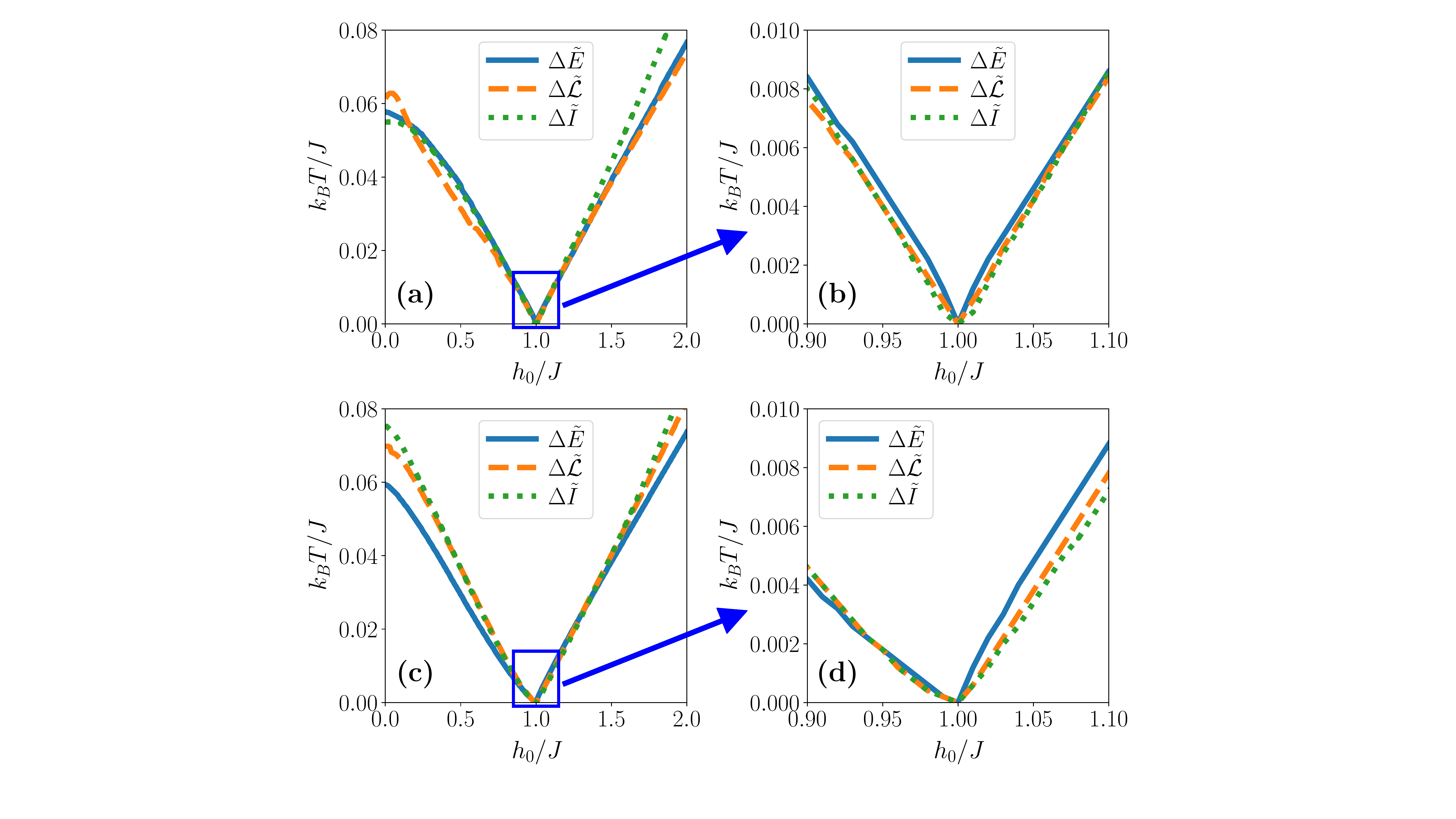}
\caption{(Color online.) Quantum critical regions in the $(h_0/J,k_BT/J)$-plane corresponding to the Ising transition (upper row) and multicritical transition (bottom row). It is constructed by solving Eq. \eqref{eq:revealqcr-macro} with energy absorbed, logarithmic negativity and quantum mutual information. For (a) and (c), the temperature $k_BT/J \lesssim 0.1$, while in (b) and (d) $k_BT/J$  is plotted upto $0.01$. All the axes are dimensionless.}
\label{fig:mutualinfo}
\end{figure}

\section{Identifying  Quantum Critical Region via dynamical quantifiers}
\label{sec:marking}

We can now qualitatively discern amongst the different regimes of the equilibrium phase diagram viz. ordered, disordered and quantum critical regions by the distinct temperature dependence of the the macroscopic and microscopic quantifiers,  $\Delta \tilde{\mathcal{Q}}$, for different quantifiers $\mathcal{Q}$. With this qualitative success, it can be expected that  it would also be possible to estimate the extent of the 
quantum critical regime quantitatively by studying the \emph{dynamics}. 

The main observation that we obtain till now is that 
\begin{equation}
\Delta \tilde{\mathcal{Q}}(T) \approx \Delta \tilde{\mathcal{Q}}(T=0) \,\mbox{for} \, \, 0 \leq T \leq T^*,
\label{eq:constant}
\end{equation}
 where $T^*$ is fixed by the initial field strength, $h_0$. We now propose that given a quantifier, $\mathcal{Q}$, and a value of $h_0$, the boundary of the QCR can be indicated by the temperature $T^*$ upto which the constancy window lasts, i.e., Eq. (\ref{eq:constant}) is valid. $T^*$ can be computed from the solution of the following condition:
%
\begin{eqnarray}
 \frac{|\Delta \mathcal{Q}_{\max}(T) - \Delta \mathcal{Q}_{\max}(T=0)|}{|\Delta \mathcal{Q}_{\max}(T=0)|} = \eta.
\label{eq:revealqcr-macro}
\end{eqnarray}
In principle, $\eta$ should be zero, but numerically, it is chosen to be as close as possible to zero. Since evaluation of $T^*$ from  Eq. \eqref{eq:revealqcr-macro} is basically a root finding problem, $\eta$  turns out to be the tolerance which is fixed by the numerical precision in our calculations. In our analysis, we fix $\eta$ to be $10^{-6}$. \emph{It means that if the fractional change in  $\Delta \tilde{\mathcal{Q}}(T)$ is below the cutoff $\eta$, it is considered to be constant, while $\Delta \tilde{\mathcal{Q}}(T) > \eta = 10^{-6} $} implies entry into the QCR. 
Let us now investigate whether such proposal works by studying the trends of $\mathcal{Q} \in \lbrace E,\mathcal{L},\mathcal{I} \rbrace$.

Fig. \ref{fig:mutualinfo} reveals the QCR obtained from three different dynamical quantifies mentioned above -- upper row corresponds to the Ising criticality, while the bottom row is for multicritical transition, which as mentioned earlier, possesses different critical exponents compared to the Ising transition. Interestingly, response of dynamical quantifiers to QCRs corresponding to these qualitatively different criticalities are almost identical. It highlights the universality of our method to compute the QCR. Furthermore, note that the QCR obtained from $\mathcal{Q} \in \lbrace E,\mathcal{L},\mathcal{I} \rbrace$, i.e from $\Delta \tilde{E}(T)$, $\Delta \tilde{\mathcal{L}}(T)$ and $\Delta \tilde{\mathcal{I}}(T)$ are qualitatively similar, i.e., the demarcated regions have a large overlap, see Figs. \ref{fig:mutualinfo}(a)-(d). 
We want to point out that the small \emph{quantitative} differences 
are not at all alarming but rather expected, since the boundary of the QCR is not at all sharp, and there is an intrinsic fuzziness. The fact that there exists no order parameter or gap closing argument as in the case of zero temperature QPTs further reinforces the above statement. In reality, the uncany similarities in the QCR as detected by different quantities, which are chosen from different paradigms, one thermodynamic and the others microscopic, are rather remarkable.

\section{Conclusion}
\label{sec:concl}
The coexistence of thermal and quantum fluctuations in the quantum critical region (QCR) makes the study of quantum phases in this regime very interesting. Although the proposal of QCR was made since the discovery of quantum phase transitions, due to the lack of any conclusive marker, very few investigations were made in this direction. Nonetheless, some attempts were made to study the QCR in equilibrium using varied techniques.

In this article, we investigated the effect of QCR on the dynamics of physical quantities after a sudden quench of the system parameters. We reported a macroscopic quantity, namely the energy absorbed during a time pulse, and two microscopic ones from the information theoretic domain - entanglement (a measure of quantum correlations) and quantum mutual information (a measure of total correlations), which provide well defined and qualitatively similar response to the presence or absence of the QCR. We quantified their responses, and converted them into detection criterion for obtaining the boundary of QCR. These quantities, which lie completely on opposite ends of the spectrum distinguish the QCR from other phases efficiently.  We want to mention that our techniques of QCR detection would suffer from computability issues when one considers systems having no analytical solutions due to the presence of both temperature and time. Nevertheless, we expect similar features for non-integrable models for finite time scales, in particular, for times much shorter than the time scales required for thermalization.  
 
We believe, our work can shed some light on the understanding of the coveted QCR. The proposal of dynamical quantifiers is relevant  from the perspective of  experiments, and at the same time,   they provide fundamental insights into the dynamical behavior of quantum critical matter -- both can have important consequence in manufacturing quantum technologies.

%
%
%

\acknowledgements
We acknowledge the support from Interdisciplinary Cyber Physical Systems (ICPS) program of the Department of Science and Technology (DST), India, Grant No.: DST/ICPS/QuST/Theme- 1/2019/23.
TC  was  supported  by Quantera  QTFLAG  project  2017/25/Z/ST2/03029  of National Science Centre (Poland).

\appendix
\section{Diagonalization of the model}
\label{Appendix}
We consider a quantum $XY$ model on a one dimensional (1D) lattice with $N$ sites, described by a Hamiltonian, 

\begin{eqnarray}
&\hat{H}&= \frac{1}{2}\sum_{j = 1}^{N}\Big[J\Big\{\frac{1+\gamma}{2}\hat{\sigma}_j^x\hat{\sigma}_{j+1}^{x}+\frac{1-\gamma}{2}\hat{\sigma}_j^y\hat{\sigma}_{j+1}^{y}\Big\} 
+h(t)\hat{\sigma}_j^z\Big] \nonumber \\
\label{ham_spin}
\end{eqnarray}
To diagonalize the $XY$ model, it is mapped to a free Fermi gas Hamiltonian by the Jordan-Wigner transformation in terms of the fermionic operator $\hat{c}$ \cite{lsm,bm1,bm2,bm3}, as 
\begin{eqnarray}
\hat{H} &=& \frac{J}{2}\sum_{j = 1}^N \Big[ \hat{c}_j^\dagger \hat{c}_{j+1}  +  \hat{c}_{j+1}^\dagger \hat{c}_j
\nonumber \\ 
 &+& \gamma  (\hat{c}_j^\dagger \hat{c}_{j+1}^\dagger + \hat{c}_{j+1} \hat{c}_{j})
+   \frac{h(t)}{J}\big( 2 \hat{c}_j^\dagger  \hat{c}_j  - 1\big)\Big].
\label{H_fermi_uniform}
\end{eqnarray}
Further a Fourier transformation enables us to write
 $\hat{H} = \sum_{p} \hat{H}_p$, where the matrix form of $\hat{H}_p$ in the basis $\lbrace|0\rangle, \hat{c}_p^\dagger \hat{c}_{-p}^\dagger |0\rangle, \hat{c}_p^\dagger  |0\rangle, \hat{c}_{-p}^\dagger |0\rangle\ \rbrace$, is given by
\begin{equation}
\resizebox{.95\hsize}{!}{$\hat{H}_p = J \begin{bmatrix}
    -h(t)/J & i\gamma \sin\phi_p & 0 & 0  \\
    - i\gamma \sin\phi_p & h(t)/J + 2\cos\phi_p & 0 & 0 \\
    0 & 0 & \cos\phi_p & 0 \\
    0 & 0 & 0 & \cos\phi_p
  \end{bmatrix},$}
\label{eq:hp_matrix_uni}
\end{equation} 
where $\phi_p = 2\pi  p /N$.

Using this simplified form of the block diagonal Hamiltonian, we can easily compute the two-site nearest-neighbor reduced density matrix between $j^{\text{th}}$ and $(j+1)^{\text{th}}$ lattice sites of the thermal state (at a temperature $T$) in equilibrium as
\begin{eqnarray}
\hat{\rho}_{j,j+1} (T) &=& \frac{1}{4} \Big[ \mathbb{I}_j \otimes \mathbb{I}_{j+1} + m^z(T) \big(\hat{\sigma}^z_j \otimes \mathbb{I}_{j+1} + \mathbb{I}_j \otimes \hat{\sigma}_{j+1}^z \big) \nonumber \\
&+& \sum_{\alpha,\alpha^\prime = x,y,z} C^{\alpha \alpha^\prime}(T) \hat{\sigma}_{j}^{\alpha} \otimes \hat{\sigma}_{j+1}^{\alpha^\prime} \Big],
\label{rho2}
\end{eqnarray}
where $m^z(T) = \langle \hat{\sigma}^z_j \otimes \mathbb{I}_{j+1} \rangle_T = \langle\mathbb{I}_j \otimes \hat{\sigma}_{j+1}^z \rangle_T$ is the transverse magnetization (along the $z$-direction) and $C^{\alpha \alpha'}(T) = \langle \hat{\sigma}_j^{\alpha} \otimes \hat{\sigma}_{j+1}^{\alpha^\prime} \rangle_T$ (with $\alpha$ and $\alpha'$ taking values $x$, $y$ or $z$ independently) are the classical correlators. Note that for any thermal state of the $XY$ model, $m^x, m^y$ and off diagonal classical correlators are identically zero.  The translation invariance of the model assures all two site reduced density matrices to be identical, and consequently all the classical correlators and magnetizations to be site independent. Thus, we drop the site labels and call the nearest neighbor density matrix to be $\hat{\rho}_{AB}$.  
However, under sudden quenches,
\begin{eqnarray}
h(t)= \left\{
 \begin{array}{cc}
 h_0, & t\leq 0  \\
 h_1, & t>0
\end{array}\right..
\end{eqnarray}
$C^{xy}$ and $C^{yx}$ also take finite values. The time evolved two site reduced density matrix, $\hat{\rho}_{AB}(t,T)$,  can be computed by tracking the time evolution of the classical correlators and magnetization, which reads as
\begin{eqnarray}
\hat{\rho}_{AB}(t,T) &=& \frac{1}{4} \Big[ \mathbb{I}_A \otimes \mathbb{I}_{B} + m^z(t,T) \big(\hat{\sigma}^z_A \otimes \mathbb{I}_{B} + \mathbb{I}_A \otimes \hat{\sigma}_{A}^z \big) \nonumber \\
&+& \sum_{\alpha = x,y,z} C^{\alpha \alpha}(t,T) \hat{\sigma}_{A}^{\alpha} \otimes \hat{\sigma}_{B}^{\alpha} \nonumber \\
&+& C^{xy}(t,T) \hat{\sigma}_A^{x} \otimes \hat{\sigma}_{B}^{y} + C^{yx}(t,T) \hat{\sigma}_A^{y} \otimes \hat{\sigma}_{B}^{x} \Big]. \nonumber \\
\label{eq:rho-t}
\end{eqnarray} 
The time dependence of these quantities $m^z(t,T)$ and the classical correlators$\big)$ after a sudden quench is analytically computed in \cite{bm1,bm2,bm3}. Furthermore, for the $XY$ model, we have $C^{xy}(t,T) = C^{yx}(t,T)$ for all times.

\subsection{Evaluation of energy absorbed in a time pulse}

The energy absorbed in a square time pulse that last a time $\tau$ is evaluated as
\begin{eqnarray}
\Delta E(T, \tau) = \text{Tr}(\hat{H}_0 e^{-i \hat{H}_1 \tau} \rho_0 e^{i \hat{H}_1 \tau}) - \text{Tr}(\hat{H}_0 \rho_0)
\end{eqnarray} 
where $\rho_0 = e^{-\frac{\hat{H}_0}{k_B T}}/\text{Tr }e^{-\frac{\hat{H}_0}{k_B T}}$. Now, after Fourier transformation, we have
\begin{eqnarray}
\Delta E(T,\tau)=\frac{1}{2\pi}\int dp \big[\text{Tr}(\hat{H}_0^p e^{-i \hat{H}_1^p \tau} \rho_0^p e^{i \hat{H}_1^p \tau}) - \text{Tr}(\hat{H}_0^p \rho_0^p)\big], \nonumber \\
\end{eqnarray}
where, $\hat{H}_{0,1}^p$ is obtained from Eq. \eqref{eq:hp_matrix_uni}.  This enormously simplifies the calculation since operators in each momentum block are just $4\times 4$ matrices.
The exponentials can be computed by diagonalizing the operators in the exponent by appropriate unitary operation, which leads to the following expression for  $E(T,\tau)$:
\small
\begin{eqnarray}
E(T,\tau)=\frac{1}{2\pi}\int dp \big[\text{Tr}(\mathcal{U} \tilde{H}_0^p \mathcal{U}^\dagger e^{-i \tilde{H}_1^p \tau} \mathcal{U} \tilde{\rho}_0^p \mathcal{U}^\dagger e^{i \tilde{H}_1^p \tau})- \text{Tr}(\tilde{H}_0^p \tilde{\rho}_0^p)\big], \nonumber \\
\end{eqnarray}
\normalsize
where we have
\begin{eqnarray}
&\tilde{H}&_{0,1}^p = \text{ diag}(\cos\phi_p - \Lambda_{0,1},\cos\phi_p + \Lambda_{0,1},\cos\phi_p,\cos\phi_p), \nonumber \\
&e&^{\pm i \tilde{H}_1^p\tau} = e^{\pm i \tau \cos \phi_p} \text{ diag}(e^{\mp i \tau \Lambda_{1}}, e^{\pm i \tau  \Lambda_{1}}, 1,  1), \nonumber \\
&\Lambda&_{0,1} = \sqrt{(\cos \phi_p + h_{0,1})^2 + \gamma^2\sin^2 \phi_p}.  
\end{eqnarray} 

%
%
Furthermore, the unitary $\mathcal{U} = U_1^\dagger U_0  \oplus \mathbb{I}_2$, where $U_{0,1}$ reads as
\begin{equation}
 U_{0,1}= \frac{1}{\sqrt{2\Lambda_{0,1}}} 
\begin{bmatrix}
  i\sqrt{\Lambda_{0,1} + \alpha_{0,1}}      & i\sqrt{\Lambda_{0,1} - \alpha_{0,1}} \\
      -\sqrt{\Lambda_{0,1} - \alpha_{0,1}}    &  \sqrt{\Lambda_{0,1} + \alpha_{0,1}}
\end{bmatrix},
\label{eq:U01}
\end{equation} 
with $\alpha_{0,1} = h_{0,1} + \cos\phi_p$. Note that $U_{0,1}  \oplus \mathbb{I}_2$ is the unitary which diagonalizes $\hat{H}_{0,1}^p$. Once we get $\tilde{H}_{0}^p$, $\tilde{\rho}_{0}^p$ can be written as $e^{-\beta \tilde{H}_{0}^p}/$Tr $e^{-\beta \tilde{H}_{0}^p}$, and can be computed easily.

\subsection{Evaluation of entanglement}

For $\hat{\rho}_{AB}(t,T)$, we always get a state in which the only non-zero local magnetization is in the z-direction, $m^z(t,T)$  and  possess the following correlation matrix 
\begin{eqnarray}
\mathcal{T}(t,T) = 
\begin{bmatrix}
  C^{xx}(t,T)      & C^{xy}(t,T) & 0 \\
      C^{yx}(t,T)    &  C^{yy}(t,T) & 0 \\
      0 & 0 & C^{zz}(t,T)
\end{bmatrix} ,
\end{eqnarray}
with $C^{xy}(t,T)=C^{yx}(t,T)$. The computation of entanglement simplifies with the observation that $\hat{\rho}_{AB}(t,T)$ can be 
brought in the standard $X$-state \cite{x1,x2} form (diagonal correlation matrix with non-zero magnetization in the $z$-direction) 
by choosing appropriate  unitaries in the local $x-y$ sectors. It is equivalent to the diagonalization of the upper $2 \times 2$ 
block of $\mathcal{T}(t,T)$. Since we have $m^x(t,T)$ and $m^y(t,T)$ to be identically $0$, they would remain so for all bases in the 
$x-y$ plane. Furthermore,  basis transformation in the local $x-y$ planes would keep $m^z(t,T)$ and $C^{zz}(t,T)$ unaltered. Since 
entanglement remains unchanged by local unitary operations, we compute the same for $\hat{\rho}_{AB}(t,T)$ using negativity \cite{negativity} , and it reads as
\begin{eqnarray}
\mathcal{N}(\hat{\rho}_{AB}(t,T)) =  -\frac{1}{4} \min \Big[  0, \mathcal{N}_1, \mathcal{N}_2 \Big],  
 \end{eqnarray}
where
\begin{eqnarray}
\mathcal{N}_1 = 1 &+& C^{zz}(t,T) \nonumber \\ &-& \sqrt{\big(C^{xx}(t,T)+C^{yy}(t,T)\big)^2 + 4(m^z(t,T))^2}, \nonumber \\
\mathcal{N}_2 = 1 &-& C^{zz}(t,T) \nonumber \\ &-& \sqrt{\big(C^{xx}(t,T)-C^{yy}(t,T)\big)^2 + 4\big(C^{xy}(t,T)\big)^2}. \nonumber \\
\end{eqnarray}
Logarithmic negativity \cite{ln} is then easily calculated to be $\mathcal{L}(t,T)=\log_2 (2\mathcal{N}(\hat{\rho}_{AB}(t,T)) +1)$.

\subsection{Evaluation of mutual information}

For computing the mutual information, we first evaluate the spectrum of $\hat{\rho}_{AB}(t,T)$, and its single site reductions, $\hat{\rho}_{A}(t,T)=$ Tr$_{B}\hat{\rho}_{AB}(t,T)$ which is equal to $\hat{\rho}_{B}(t,T)=$ Tr$_{A}\hat{\rho}_{AB}(t,T)$. The spectrum of $\hat{\rho}_{AB}(t,T)$, $\mathcal{X}_{\hat{\rho}_{AB}(t,T)}$, is obtained from the eigenvalues of $\hat{\rho}_{AB}(t,T)$ and is given by
\begin{eqnarray}
&\mathcal{X}&_{\hat{\rho}_{AB}(t,T)}  = \frac{1}{4} \times \nonumber \\
&\Big \lbrace& 1 - C^{zz}(t,T) \pm \vert C^{xx}(t,T) + C^{yy}(t,T) \vert, 
 1 + C^{zz}(t,T) \nonumber \\ 
 &\pm& \sqrt{(C^{xx}(t,T) - C^{yy}(t,T))^2 + 4 (m^z(t,T))^2} \Big \rbrace. 
\end{eqnarray} 
The spectra of $\hat{\rho}_{A(B)}(t,T)$ is computed to be
\begin{eqnarray}
\mathcal{X}_{\hat{\rho}_{A}(t,T)} = \mathcal{X}_{\hat{\rho}_{B}(t,T)} = \frac{1}{2}\big \lbrace 1 \pm m^z(t,T) \big \rbrace.
\end{eqnarray}
Therefore, the mutual information between $A$ and $B$ of $\hat{\rho}_{AB}(t)$ can be expressed as
\begin{eqnarray}
\mathcal{I}_{A:B}(t,T) = H(\mathcal{X}_{\hat{\rho}_{A}(t,T)}) + H(\mathcal{X}_{\hat{\rho}_{B}(t,T)}) - H(\mathcal{X}_{\hat{\rho}_{AB}(t,T)}), \nonumber \\
\end{eqnarray}
where $H(\mathcal{X})$ denotes the Shannon entropy of the spectral (probability) distribution $\mathcal{X}$. Since we have $\mathcal{X}_{\hat{\rho}_{A}(t,T)} = \mathcal{X}_{\hat{\rho}_{B}(t,T)}$, the above equation can be simplified as
\begin{eqnarray}
\mathcal{I}_{A:B}(t,T) = 2H(\mathcal{X}_{\hat{\rho}_{A}(t,T)}) - H(\mathcal{X}_{\hat{\rho}_{AB}(t,T)}).
\end{eqnarray}

\vspace{0.25cm}

\bibliography{bib}

\begin{thebibliography}{50}%
\makeatletter
\providecommand \@ifxundefined [1]{%
 \@ifx{#1\undefined}
}%
\providecommand \@ifnum [1]{%
 \ifnum #1\expandafter \@firstoftwo
 \else \expandafter \@secondoftwo
 \fi
}%
\providecommand \@ifx [1]{%
 \ifx #1\expandafter \@firstoftwo
 \else \expandafter \@secondoftwo
 \fi
}%
\providecommand \natexlab [1]{#1}%
\providecommand \enquote  [1]{``#1''}%
\providecommand \bibnamefont  [1]{#1}%
\providecommand \bibfnamefont [1]{#1}%
\providecommand \citenamefont [1]{#1}%
\providecommand \href@noop [0]{\@secondoftwo}%
\providecommand \href [0]{\begingroup \@sanitize@url \@href}%
\providecommand \@href[1]{\@@startlink{#1}\@@href}%
\providecommand \@@href[1]{\endgroup#1\@@endlink}%
\providecommand \@sanitize@url [0]{\catcode `\\12\catcode `\$12\catcode
  `\&12\catcode `\#12\catcode `\^12\catcode `\_12\catcode `\%12\relax}%
\providecommand \@@startlink[1]{}%
\providecommand \@@endlink[0]{}%
\providecommand \url  [0]{\begingroup\@sanitize@url \@url }%
\providecommand \@url [1]{\endgroup\@href {#1}{\urlprefix }}%
\providecommand \urlprefix  [0]{URL }%
\providecommand \Eprint [0]{\href }%
\providecommand \doibase [0]{http://dx.doi.org/}%
\providecommand \selectlanguage [0]{\@gobble}%
\providecommand \bibinfo  [0]{\@secondoftwo}%
\providecommand \bibfield  [0]{\@secondoftwo}%
\providecommand \translation [1]{[#1]}%
\providecommand \BibitemOpen [0]{}%
\providecommand \bibitemStop [0]{}%
\providecommand \bibitemNoStop [0]{.\EOS\space}%
\providecommand \EOS [0]{\spacefactor3000\relax}%
\providecommand \BibitemShut  [1]{\csname bibitem#1\endcsname}%
\let\auto@bib@innerbib\@empty
\bibitem [{\citenamefont {Chakrabarti}\ \emph {et~al.}(1996)\citenamefont
  {Chakrabarti}, \citenamefont {Dutta},\ and\ \citenamefont {Sen}}]{qptbook4}%
  \BibitemOpen
  \bibfield  {author} {\bibinfo {author} {\bibfnamefont {B.~K.}\ \bibnamefont
  {Chakrabarti}}, \bibinfo {author} {\bibfnamefont {A.}~\bibnamefont {Dutta}},
  \ and\ \bibinfo {author} {\bibfnamefont {P.}~\bibnamefont {Sen}},\ }\href
  {\doibase 10.1007/978-3-540-49865-0} {\emph {\bibinfo {title} {Quantum Ising
  phases and transitions in transverse Ising models}}}\ (\bibinfo  {publisher}
  {Springer Berlin Heidelberg},\ \bibinfo {year} {1996})\BibitemShut {NoStop}%
\bibitem [{\citenamefont {Sachdev}(2009)}]{qptbook2}%
  \BibitemOpen
  \bibfield  {author} {\bibinfo {author} {\bibfnamefont {S.}~\bibnamefont
  {Sachdev}},\ }\href {\doibase 10.1017/cbo9780511973765} {\emph {\bibinfo
  {title} {Quantum Phase Transitions}}}\ (\bibinfo  {publisher} {Cambridge
  University Press},\ \bibinfo {year} {2009})\BibitemShut {NoStop}%
\bibitem [{\citenamefont {Suzuki}\ \emph {et~al.}(2013)\citenamefont {Suzuki},
  \citenamefont {Inoue},\ and\ \citenamefont {Chakrabarti}}]{qptbook3}%
  \BibitemOpen
  \bibfield  {author} {\bibinfo {author} {\bibfnamefont {S.}~\bibnamefont
  {Suzuki}}, \bibinfo {author} {\bibfnamefont {J.}~\bibnamefont {Inoue}}, \
  and\ \bibinfo {author} {\bibfnamefont {B.~K.}\ \bibnamefont {Chakrabarti}},\
  }\href {\doibase 10.1007/978-3-642-33039-1} {\emph {\bibinfo {title} {Quantum
  Ising Phases and Transitions in Transverse Ising Models}}}\ (\bibinfo
  {publisher} {Springer Berlin Heidelberg},\ \bibinfo {year}
  {2013})\BibitemShut {NoStop}%
\bibitem [{\citenamefont {Dutta}\ \emph {et~al.}(2015)\citenamefont {Dutta},
  \citenamefont {Aeppli}, \citenamefont {Chakrabarti}, \citenamefont
  {Divakaran}, \citenamefont {Rosenbaum},\ and\ \citenamefont
  {Sen}}]{qptbook1}%
  \BibitemOpen
  \bibfield  {author} {\bibinfo {author} {\bibfnamefont {A.}~\bibnamefont
  {Dutta}}, \bibinfo {author} {\bibfnamefont {G.}~\bibnamefont {Aeppli}},
  \bibinfo {author} {\bibfnamefont {B.~K.}\ \bibnamefont {Chakrabarti}},
  \bibinfo {author} {\bibfnamefont {U.}~\bibnamefont {Divakaran}}, \bibinfo
  {author} {\bibfnamefont {T.~F.}\ \bibnamefont {Rosenbaum}}, \ and\ \bibinfo
  {author} {\bibfnamefont {D.}~\bibnamefont {Sen}},\ }\href {\doibase
  10.1017/CBO9781107706057} {\emph {\bibinfo {title} {Quantum Phase Transitions
  in Transverse Field Spin Models: From Statistical Physics to Quantum
  Information}}}\ (\bibinfo  {publisher} {Cambridge University Press},\
  \bibinfo {year} {2015})\BibitemShut {NoStop}%
\bibitem [{\citenamefont {Sengupta}\ \emph {et~al.}(2004)\citenamefont
  {Sengupta}, \citenamefont {Powell},\ and\ \citenamefont {Sachdev}}]{KS2004}%
  \BibitemOpen
  \bibfield  {author} {\bibinfo {author} {\bibfnamefont {K.}~\bibnamefont
  {Sengupta}}, \bibinfo {author} {\bibfnamefont {S.}~\bibnamefont {Powell}}, \
  and\ \bibinfo {author} {\bibfnamefont {S.}~\bibnamefont {Sachdev}},\ }\href
  {\doibase 10.1103/PhysRevA.69.053616} {\bibfield  {journal} {\bibinfo
  {journal} {Phys. Rev. A}\ }\textbf {\bibinfo {volume} {69}},\ \bibinfo
  {pages} {053616} (\bibinfo {year} {2004})}\BibitemShut {NoStop}%
\bibitem [{\citenamefont {Sen(De)}\ \emph {et~al.}(2005)\citenamefont
  {Sen(De)}, \citenamefont {Sen},\ and\ \citenamefont {Lewenstein}}]{ASD2005}%
  \BibitemOpen
  \bibfield  {author} {\bibinfo {author} {\bibfnamefont {A.}~\bibnamefont
  {Sen(De)}}, \bibinfo {author} {\bibfnamefont {U.}~\bibnamefont {Sen}}, \ and\
  \bibinfo {author} {\bibfnamefont {M.}~\bibnamefont {Lewenstein}},\ }\href
  {\doibase 10.1103/PhysRevA.72.052319} {\bibfield  {journal} {\bibinfo
  {journal} {Phys. Rev. A}\ }\textbf {\bibinfo {volume} {72}},\ \bibinfo
  {pages} {052319} (\bibinfo {year} {2005})}\BibitemShut {NoStop}%
\bibitem [{\citenamefont {Pollmann}\ \emph {et~al.}(2010)\citenamefont
  {Pollmann}, \citenamefont {Mukerjee}, \citenamefont {Green},\ and\
  \citenamefont {Moore}}]{pollmann}%
  \BibitemOpen
  \bibfield  {author} {\bibinfo {author} {\bibfnamefont {F.}~\bibnamefont
  {Pollmann}}, \bibinfo {author} {\bibfnamefont {S.}~\bibnamefont {Mukerjee}},
  \bibinfo {author} {\bibfnamefont {A.~G.}\ \bibnamefont {Green}}, \ and\
  \bibinfo {author} {\bibfnamefont {J.~E.}\ \bibnamefont {Moore}},\ }\href
  {\doibase 10.1103/PhysRevE.81.020101} {\bibfield  {journal} {\bibinfo
  {journal} {Phys. Rev. E}\ }\textbf {\bibinfo {volume} {81}},\ \bibinfo
  {pages} {020101} (\bibinfo {year} {2010})}\BibitemShut {NoStop}%
\bibitem [{\citenamefont {Heyl}\ \emph {et~al.}(2013)\citenamefont {Heyl},
  \citenamefont {Polkovnikov},\ and\ \citenamefont {Kehrein}}]{heyl_prl}%
  \BibitemOpen
  \bibfield  {author} {\bibinfo {author} {\bibfnamefont {M.}~\bibnamefont
  {Heyl}}, \bibinfo {author} {\bibfnamefont {A.}~\bibnamefont {Polkovnikov}}, \
  and\ \bibinfo {author} {\bibfnamefont {S.}~\bibnamefont {Kehrein}},\ }\href
  {\doibase 10.1103/PhysRevLett.110.135704} {\bibfield  {journal} {\bibinfo
  {journal} {Phys. Rev. Lett.}\ }\textbf {\bibinfo {volume} {110}},\ \bibinfo
  {pages} {135704} (\bibinfo {year} {2013})}\BibitemShut {NoStop}%
\bibitem [{\citenamefont {Heyl}(2014)}]{heyl_prl_2}%
  \BibitemOpen
  \bibfield  {author} {\bibinfo {author} {\bibfnamefont {M.}~\bibnamefont
  {Heyl}},\ }\href {\doibase 10.1103/PhysRevLett.113.205701} {\bibfield
  {journal} {\bibinfo  {journal} {Phys. Rev. Lett.}\ }\textbf {\bibinfo
  {volume} {113}},\ \bibinfo {pages} {205701} (\bibinfo {year}
  {2014})}\BibitemShut {NoStop}%
\bibitem [{\citenamefont {\ifmmode \check{Z}\else
  \v{Z}\fi{}unkovi\ifmmode~\check{c}\else \v{c}\fi{}}\ \emph
  {et~al.}(2018)\citenamefont {\ifmmode \check{Z}\else
  \v{Z}\fi{}unkovi\ifmmode~\check{c}\else \v{c}\fi{}}, \citenamefont {Heyl},
  \citenamefont {Knap},\ and\ \citenamefont {Silva}}]{heyl_prl_3}%
  \BibitemOpen
  \bibfield  {author} {\bibinfo {author} {\bibfnamefont {B.}~\bibnamefont
  {\ifmmode \check{Z}\else \v{Z}\fi{}unkovi\ifmmode~\check{c}\else
  \v{c}\fi{}}}, \bibinfo {author} {\bibfnamefont {M.}~\bibnamefont {Heyl}},
  \bibinfo {author} {\bibfnamefont {M.}~\bibnamefont {Knap}}, \ and\ \bibinfo
  {author} {\bibfnamefont {A.}~\bibnamefont {Silva}},\ }\href {\doibase
  10.1103/PhysRevLett.120.130601} {\bibfield  {journal} {\bibinfo  {journal}
  {Phys. Rev. Lett.}\ }\textbf {\bibinfo {volume} {120}},\ \bibinfo {pages}
  {130601} (\bibinfo {year} {2018})}\BibitemShut {NoStop}%
\bibitem [{\citenamefont {Heyl}(2015)}]{heylprl4}%
  \BibitemOpen
  \bibfield  {author} {\bibinfo {author} {\bibfnamefont {M.}~\bibnamefont
  {Heyl}},\ }\href {\doibase 10.1103/PhysRevLett.115.140602} {\bibfield
  {journal} {\bibinfo  {journal} {Phys. Rev. Lett.}\ }\textbf {\bibinfo
  {volume} {115}},\ \bibinfo {pages} {140602} (\bibinfo {year}
  {2015})}\BibitemShut {NoStop}%
\bibitem [{\citenamefont {Heyl}(2018)}]{heyl_review}%
  \BibitemOpen
  \bibfield  {author} {\bibinfo {author} {\bibfnamefont {M.}~\bibnamefont
  {Heyl}},\ }\href {http://stacks.iop.org/0034-4885/81/i=5/a=054001} {\bibfield
   {journal} {\bibinfo  {journal} {Rep. Prog. Phys.}\ }\textbf {\bibinfo
  {volume} {81}},\ \bibinfo {pages} {054001} (\bibinfo {year}
  {2018})}\BibitemShut {NoStop}%
\bibitem [{\citenamefont {Weidinger}\ \emph {et~al.}(2017)\citenamefont
  {Weidinger}, \citenamefont {Heyl}, \citenamefont {Silva},\ and\ \citenamefont
  {Knap}}]{heyl_cont_sym_break}%
  \BibitemOpen
  \bibfield  {author} {\bibinfo {author} {\bibfnamefont {S.~A.}\ \bibnamefont
  {Weidinger}}, \bibinfo {author} {\bibfnamefont {M.}~\bibnamefont {Heyl}},
  \bibinfo {author} {\bibfnamefont {A.}~\bibnamefont {Silva}}, \ and\ \bibinfo
  {author} {\bibfnamefont {M.}~\bibnamefont {Knap}},\ }\href {\doibase
  10.1103/PhysRevB.96.134313} {\bibfield  {journal} {\bibinfo  {journal} {Phys.
  Rev. B}\ }\textbf {\bibinfo {volume} {96}},\ \bibinfo {pages} {134313}
  (\bibinfo {year} {2017})}\BibitemShut {NoStop}%
\bibitem [{\citenamefont {Vajna}\ and\ \citenamefont
  {D\'ora}(2014)}]{Vajna_prb}%
  \BibitemOpen
  \bibfield  {author} {\bibinfo {author} {\bibfnamefont {S.}~\bibnamefont
  {Vajna}}\ and\ \bibinfo {author} {\bibfnamefont {B.}~\bibnamefont {D\'ora}},\
  }\href {\doibase 10.1103/PhysRevB.89.161105} {\bibfield  {journal} {\bibinfo
  {journal} {Phys. Rev. B}\ }\textbf {\bibinfo {volume} {89}},\ \bibinfo
  {pages} {161105} (\bibinfo {year} {2014})}\BibitemShut {NoStop}%
\bibitem [{\citenamefont {Canovi}\ \emph {et~al.}(2014)\citenamefont {Canovi},
  \citenamefont {Ercolessi}, \citenamefont {Naldesi}, \citenamefont {Taddia},\
  and\ \citenamefont {Vodola}}]{schmidtgap}%
  \BibitemOpen
  \bibfield  {author} {\bibinfo {author} {\bibfnamefont {E.}~\bibnamefont
  {Canovi}}, \bibinfo {author} {\bibfnamefont {E.}~\bibnamefont {Ercolessi}},
  \bibinfo {author} {\bibfnamefont {P.}~\bibnamefont {Naldesi}}, \bibinfo
  {author} {\bibfnamefont {L.}~\bibnamefont {Taddia}}, \ and\ \bibinfo {author}
  {\bibfnamefont {D.}~\bibnamefont {Vodola}},\ }\href {\doibase
  10.1103/PhysRevB.89.104303} {\bibfield  {journal} {\bibinfo  {journal} {Phys.
  Rev. B}\ }\textbf {\bibinfo {volume} {89}},\ \bibinfo {pages} {104303}
  (\bibinfo {year} {2014})}\BibitemShut {NoStop}%
\bibitem [{\citenamefont {Haldar}\ \emph {et~al.}(2020)\citenamefont {Haldar},
  \citenamefont {Roy}, \citenamefont {Chanda}, \citenamefont {Sen(De)},\ and\
  \citenamefont {Sen}}]{dqpt-stav1}%
  \BibitemOpen
  \bibfield  {author} {\bibinfo {author} {\bibfnamefont {S.}~\bibnamefont
  {Haldar}}, \bibinfo {author} {\bibfnamefont {S.}~\bibnamefont {Roy}},
  \bibinfo {author} {\bibfnamefont {T.}~\bibnamefont {Chanda}}, \bibinfo
  {author} {\bibfnamefont {A.}~\bibnamefont {Sen(De)}}, \ and\ \bibinfo
  {author} {\bibfnamefont {U.}~\bibnamefont {Sen}},\ }\href {\doibase
  10.1103/PhysRevB.101.224304} {\bibfield  {journal} {\bibinfo  {journal}
  {Phys. Rev. B}\ }\textbf {\bibinfo {volume} {101}},\ \bibinfo {pages}
  {224304} (\bibinfo {year} {2020})}\BibitemShut {NoStop}%
\bibitem [{\citenamefont {Chakravarty}\ \emph {et~al.}(1989)\citenamefont
  {Chakravarty}, \citenamefont {Halperin},\ and\ \citenamefont {Nelson}}]{qd1}%
  \BibitemOpen
  \bibfield  {author} {\bibinfo {author} {\bibfnamefont {S.}~\bibnamefont
  {Chakravarty}}, \bibinfo {author} {\bibfnamefont {B.~I.}\ \bibnamefont
  {Halperin}}, \ and\ \bibinfo {author} {\bibfnamefont {D.~R.}\ \bibnamefont
  {Nelson}},\ }\href {\doibase 10.1103/PhysRevB.39.2344} {\bibfield  {journal}
  {\bibinfo  {journal} {Phys. Rev. B}\ }\textbf {\bibinfo {volume} {39}},\
  \bibinfo {pages} {2344} (\bibinfo {year} {1989})}\BibitemShut {NoStop}%
\bibitem [{\citenamefont {Sachdev}\ and\ \citenamefont {Ye}(1992)}]{qd2}%
  \BibitemOpen
  \bibfield  {author} {\bibinfo {author} {\bibfnamefont {S.}~\bibnamefont
  {Sachdev}}\ and\ \bibinfo {author} {\bibfnamefont {J.}~\bibnamefont {Ye}},\
  }\href {\doibase 10.1103/PhysRevLett.69.2411} {\bibfield  {journal} {\bibinfo
   {journal} {Phys. Rev. Lett.}\ }\textbf {\bibinfo {volume} {69}},\ \bibinfo
  {pages} {2411} (\bibinfo {year} {1992})}\BibitemShut {NoStop}%
\bibitem [{\citenamefont {Sachdev}\ and\ \citenamefont {Young}(1997)}]{qd3}%
  \BibitemOpen
  \bibfield  {author} {\bibinfo {author} {\bibfnamefont {S.}~\bibnamefont
  {Sachdev}}\ and\ \bibinfo {author} {\bibfnamefont {A.~P.}\ \bibnamefont
  {Young}},\ }\href {\doibase 10.1103/physrevlett.78.2220} {\bibfield
  {journal} {\bibinfo  {journal} {Phys. Rev. Lett.}\ }\textbf {\bibinfo
  {volume} {78}},\ \bibinfo {pages} {2220} (\bibinfo {year}
  {1997})}\BibitemShut {NoStop}%
\bibitem [{\citenamefont {Sachdev}(2000)}]{qd4}%
  \BibitemOpen
  \bibfield  {author} {\bibinfo {author} {\bibfnamefont {S.}~\bibnamefont
  {Sachdev}},\ }\href {\doibase 10.1126/science.288.5465.475} {\bibfield
  {journal} {\bibinfo  {journal} {Science}\ }\textbf {\bibinfo {volume}
  {288}},\ \bibinfo {pages} {475} (\bibinfo {year} {2000})}\BibitemShut
  {NoStop}%
\bibitem [{\citenamefont {R\"uegg}\ \emph {et~al.}(2008)\citenamefont
  {R\"uegg}, \citenamefont {Normand}, \citenamefont {Matsumoto}, \citenamefont
  {Furrer}, \citenamefont {McMorrow}, \citenamefont {Kr\"amer}, \citenamefont
  {G\"udel}, \citenamefont {Gvasaliya}, \citenamefont {Mutka},\ and\
  \citenamefont {Boehm}}]{qd5}%
  \BibitemOpen
  \bibfield  {author} {\bibinfo {author} {\bibfnamefont {C.}~\bibnamefont
  {R\"uegg}}, \bibinfo {author} {\bibfnamefont {B.}~\bibnamefont {Normand}},
  \bibinfo {author} {\bibfnamefont {M.}~\bibnamefont {Matsumoto}}, \bibinfo
  {author} {\bibfnamefont {A.}~\bibnamefont {Furrer}}, \bibinfo {author}
  {\bibfnamefont {D.~F.}\ \bibnamefont {McMorrow}}, \bibinfo {author}
  {\bibfnamefont {K.~W.}\ \bibnamefont {Kr\"amer}}, \bibinfo {author}
  {\bibfnamefont {H.~U.}\ \bibnamefont {G\"udel}}, \bibinfo {author}
  {\bibfnamefont {S.~N.}\ \bibnamefont {Gvasaliya}}, \bibinfo {author}
  {\bibfnamefont {H.}~\bibnamefont {Mutka}}, \ and\ \bibinfo {author}
  {\bibfnamefont {M.}~\bibnamefont {Boehm}},\ }\href {\doibase
  10.1103/PhysRevLett.100.205701} {\bibfield  {journal} {\bibinfo  {journal}
  {Phys. Rev. Lett.}\ }\textbf {\bibinfo {volume} {100}},\ \bibinfo {pages}
  {205701} (\bibinfo {year} {2008})}\BibitemShut {NoStop}%
\bibitem [{\citenamefont {Zhu}\ \emph {et~al.}(2015)\citenamefont {Zhu},
  \citenamefont {Wang}, \citenamefont {Sun}, \citenamefont {Xiong},
  \citenamefont {Xiong},\ and\ \citenamefont {L\"{u}}}]{qd6}%
  \BibitemOpen
  \bibfield  {author} {\bibinfo {author} {\bibfnamefont {Q.}~\bibnamefont
  {Zhu}}, \bibinfo {author} {\bibfnamefont {B.}~\bibnamefont {Wang}}, \bibinfo
  {author} {\bibfnamefont {C.}~\bibnamefont {Sun}}, \bibinfo {author}
  {\bibfnamefont {D.}~\bibnamefont {Xiong}}, \bibinfo {author} {\bibfnamefont
  {H.}~\bibnamefont {Xiong}}, \ and\ \bibinfo {author} {\bibfnamefont
  {B.}~\bibnamefont {L\"{u}}},\ }\href {\doibase 10.1088/1367-2630/17/6/063015}
  {\bibfield  {journal} {\bibinfo  {journal} {New Journal of Physics}\ }\textbf
  {\bibinfo {volume} {17}},\ \bibinfo {pages} {063015} (\bibinfo {year}
  {2015})}\BibitemShut {NoStop}%
\bibitem [{\citenamefont {feng Yang}\ \emph {et~al.}(2017)\citenamefont {feng
  Yang}, \citenamefont {Pines},\ and\ \citenamefont {Lonzarich}}]{qd7}%
  \BibitemOpen
  \bibfield  {author} {\bibinfo {author} {\bibfnamefont {Y.}~\bibnamefont {feng
  Yang}}, \bibinfo {author} {\bibfnamefont {D.}~\bibnamefont {Pines}}, \ and\
  \bibinfo {author} {\bibfnamefont {G.}~\bibnamefont {Lonzarich}},\ }\href
  {\doibase 10.1073/pnas.1703172114} {\bibfield  {journal} {\bibinfo  {journal}
  {Proceedings of the National Academy of Sciences}\ }\textbf {\bibinfo
  {volume} {114}},\ \bibinfo {pages} {6250} (\bibinfo {year}
  {2017})}\BibitemShut {NoStop}%
\bibitem [{\citenamefont {Fr{\'{e}}rot}\ and\ \citenamefont
  {Roscilde}(2019)}]{qd8}%
  \BibitemOpen
  \bibfield  {author} {\bibinfo {author} {\bibfnamefont {I.}~\bibnamefont
  {Fr{\'{e}}rot}}\ and\ \bibinfo {author} {\bibfnamefont {T.}~\bibnamefont
  {Roscilde}},\ }\href {\doibase 10.1038/s41467-019-08324-9} {\ \textbf
  {\bibinfo {volume} {10}},\ \bibinfo {pages} {577} (\bibinfo {year}
  {2019})}\BibitemShut {NoStop}%
\bibitem [{\citenamefont {Amico}\ and\ \citenamefont
  {Patan{\`{e}}}(2007)}]{Amico2008}%
  \BibitemOpen
  \bibfield  {author} {\bibinfo {author} {\bibfnamefont {L.}~\bibnamefont
  {Amico}}\ and\ \bibinfo {author} {\bibfnamefont {D.}~\bibnamefont
  {Patan{\`{e}}}},\ }\href {\doibase 10.1209/0295-5075/77/17001} {\bibfield
  {journal} {\bibinfo  {journal} {Europhysics Letters ({EPL})}\ }\textbf
  {\bibinfo {volume} {77}},\ \bibinfo {pages} {17001} (\bibinfo {year}
  {2007})}\BibitemShut {NoStop}%
\bibitem [{\citenamefont {Rane}\ \emph {et~al.}(2014)\citenamefont {Rane},
  \citenamefont {Mishra}, \citenamefont {Biswas}, \citenamefont {Sen(De)},\
  and\ \citenamefont {Sen}}]{AmeyaPRE}%
  \BibitemOpen
  \bibfield  {author} {\bibinfo {author} {\bibfnamefont {A.~D.}\ \bibnamefont
  {Rane}}, \bibinfo {author} {\bibfnamefont {U.}~\bibnamefont {Mishra}},
  \bibinfo {author} {\bibfnamefont {A.}~\bibnamefont {Biswas}}, \bibinfo
  {author} {\bibfnamefont {A.}~\bibnamefont {Sen(De)}}, \ and\ \bibinfo
  {author} {\bibfnamefont {U.}~\bibnamefont {Sen}},\ }\href {\doibase
  10.1103/PhysRevE.90.022144} {\bibfield  {journal} {\bibinfo  {journal} {Phys.
  Rev. E}\ }\textbf {\bibinfo {volume} {90}},\ \bibinfo {pages} {022144}
  (\bibinfo {year} {2014})}\BibitemShut {NoStop}%
\bibitem [{\citenamefont {Kinross}\ \emph {et~al.}(2014)\citenamefont
  {Kinross}, \citenamefont {Fu}, \citenamefont {Munsie}, \citenamefont
  {Dabkowska}, \citenamefont {Luke}, \citenamefont {Sachdev},\ and\
  \citenamefont {Imai}}]{sachdevPRX}%
  \BibitemOpen
  \bibfield  {author} {\bibinfo {author} {\bibfnamefont {A.~W.}\ \bibnamefont
  {Kinross}}, \bibinfo {author} {\bibfnamefont {M.}~\bibnamefont {Fu}},
  \bibinfo {author} {\bibfnamefont {T.~J.}\ \bibnamefont {Munsie}}, \bibinfo
  {author} {\bibfnamefont {H.~A.}\ \bibnamefont {Dabkowska}}, \bibinfo {author}
  {\bibfnamefont {G.~M.}\ \bibnamefont {Luke}}, \bibinfo {author}
  {\bibfnamefont {S.}~\bibnamefont {Sachdev}}, \ and\ \bibinfo {author}
  {\bibfnamefont {T.}~\bibnamefont {Imai}},\ }\href {\doibase
  10.1103/PhysRevX.4.031008} {\bibfield  {journal} {\bibinfo  {journal} {Phys.
  Rev. X}\ }\textbf {\bibinfo {volume} {4}},\ \bibinfo {pages} {031008}
  (\bibinfo {year} {2014})}\BibitemShut {NoStop}%
\bibitem [{\citenamefont {Barouch}\ \emph {et~al.}(1970)\citenamefont
  {Barouch}, \citenamefont {McCoy},\ and\ \citenamefont {Dresden}}]{bm1}%
  \BibitemOpen
  \bibfield  {author} {\bibinfo {author} {\bibfnamefont {E.}~\bibnamefont
  {Barouch}}, \bibinfo {author} {\bibfnamefont {B.~M.}\ \bibnamefont {McCoy}},
  \ and\ \bibinfo {author} {\bibfnamefont {M.}~\bibnamefont {Dresden}},\ }\href
  {\doibase 10.1103/PhysRevA.2.1075} {\bibfield  {journal} {\bibinfo  {journal}
  {Phys. Rev. A}\ }\textbf {\bibinfo {volume} {2}},\ \bibinfo {pages} {1075}
  (\bibinfo {year} {1970})}\BibitemShut {NoStop}%
\bibitem [{\citenamefont {Bhattacharyya}\ \emph {et~al.}(2015)\citenamefont
  {Bhattacharyya}, \citenamefont {Dasgupta},\ and\ \citenamefont
  {Das}}]{Subinoy}%
  \BibitemOpen
  \bibfield  {author} {\bibinfo {author} {\bibfnamefont {S.}~\bibnamefont
  {Bhattacharyya}}, \bibinfo {author} {\bibfnamefont {S.}~\bibnamefont
  {Dasgupta}}, \ and\ \bibinfo {author} {\bibfnamefont {A.}~\bibnamefont
  {Das}},\ }\href {\doibase 10.1038/srep16490} {\bibfield  {journal} {\bibinfo
  {journal} {Scientific Reports}\ }\textbf {\bibinfo {volume} {5}},\ \bibinfo
  {pages} {16490} (\bibinfo {year} {2015})}\BibitemShut {NoStop}%
\bibitem [{\citenamefont {Horodecki}\ \emph {et~al.}(2009)\citenamefont
  {Horodecki}, \citenamefont {Horodecki}, \citenamefont {Horodecki},\ and\
  \citenamefont {Horodecki}}]{hhhh}%
  \BibitemOpen
  \bibfield  {author} {\bibinfo {author} {\bibfnamefont {R.}~\bibnamefont
  {Horodecki}}, \bibinfo {author} {\bibfnamefont {P.}~\bibnamefont
  {Horodecki}}, \bibinfo {author} {\bibfnamefont {M.}~\bibnamefont
  {Horodecki}}, \ and\ \bibinfo {author} {\bibfnamefont {K.}~\bibnamefont
  {Horodecki}},\ }\href {\doibase 10.1103/RevModPhys.81.865} {\bibfield
  {journal} {\bibinfo  {journal} {Rev. Mod. Phys.}\ }\textbf {\bibinfo {volume}
  {81}},\ \bibinfo {pages} {865} (\bibinfo {year} {2009})}\BibitemShut
  {NoStop}%
\bibitem [{\citenamefont {Osterloh}\ \emph {et~al.}(2002)\citenamefont
  {Osterloh}, \citenamefont {Amico}, \citenamefont {Falci},\ and\ \citenamefont
  {Fazio}}]{Fazioqpt}%
  \BibitemOpen
  \bibfield  {author} {\bibinfo {author} {\bibfnamefont {A.}~\bibnamefont
  {Osterloh}}, \bibinfo {author} {\bibfnamefont {L.}~\bibnamefont {Amico}},
  \bibinfo {author} {\bibfnamefont {G.}~\bibnamefont {Falci}}, \ and\ \bibinfo
  {author} {\bibfnamefont {R.}~\bibnamefont {Fazio}},\ }\href {\doibase
  10.1038/416608a} {\bibfield  {journal} {\bibinfo  {journal} {Nature}\
  }\textbf {\bibinfo {volume} {416}},\ \bibinfo {pages} {608} (\bibinfo {year}
  {2002})}\BibitemShut {NoStop}%
\bibitem [{\citenamefont {Osborne}\ and\ \citenamefont
  {Nielsen}(2002)}]{Osborne2002}%
  \BibitemOpen
  \bibfield  {author} {\bibinfo {author} {\bibfnamefont {T.~J.}\ \bibnamefont
  {Osborne}}\ and\ \bibinfo {author} {\bibfnamefont {M.~A.}\ \bibnamefont
  {Nielsen}},\ }\href {\doibase 10.1103/PhysRevA.66.032110} {\bibfield
  {journal} {\bibinfo  {journal} {Phys. Rev. A}\ }\textbf {\bibinfo {volume}
  {66}},\ \bibinfo {pages} {032110} (\bibinfo {year} {2002})}\BibitemShut
  {NoStop}%
\bibitem [{\citenamefont {Amico}\ \emph {et~al.}(2008)\citenamefont {Amico},
  \citenamefont {Fazio}, \citenamefont {Osterloh},\ and\ \citenamefont
  {Vedral}}]{rmp-amico}%
  \BibitemOpen
  \bibfield  {author} {\bibinfo {author} {\bibfnamefont {L.}~\bibnamefont
  {Amico}}, \bibinfo {author} {\bibfnamefont {R.}~\bibnamefont {Fazio}},
  \bibinfo {author} {\bibfnamefont {A.}~\bibnamefont {Osterloh}}, \ and\
  \bibinfo {author} {\bibfnamefont {V.}~\bibnamefont {Vedral}},\ }\href
  {\doibase 10.1103/RevModPhys.80.517} {\bibfield  {journal} {\bibinfo
  {journal} {Rev. Mod. Phys.}\ }\textbf {\bibinfo {volume} {80}},\ \bibinfo
  {pages} {517} (\bibinfo {year} {2008})}\BibitemShut {NoStop}%
\bibitem [{\citenamefont {Lewenstein}\ \emph {et~al.}(2007)\citenamefont
  {Lewenstein}, \citenamefont {Sanpera}, \citenamefont {Ahufinger},
  \citenamefont {Damski}, \citenamefont {Sen(De)},\ and\ \citenamefont
  {Sen}}]{aditi-mac}%
  \BibitemOpen
  \bibfield  {author} {\bibinfo {author} {\bibfnamefont {M.}~\bibnamefont
  {Lewenstein}}, \bibinfo {author} {\bibfnamefont {A.}~\bibnamefont {Sanpera}},
  \bibinfo {author} {\bibfnamefont {V.}~\bibnamefont {Ahufinger}}, \bibinfo
  {author} {\bibfnamefont {B.}~\bibnamefont {Damski}}, \bibinfo {author}
  {\bibfnamefont {A.}~\bibnamefont {Sen(De)}}, \ and\ \bibinfo {author}
  {\bibfnamefont {U.}~\bibnamefont {Sen}},\ }\href {\doibase
  10.1080/00018730701223200} {\bibfield  {journal} {\bibinfo  {journal}
  {Advances in Physics}\ }\textbf {\bibinfo {volume} {56}},\ \bibinfo {pages}
  {243} (\bibinfo {year} {2007})}\BibitemShut {NoStop}%
\bibitem [{\citenamefont {Lewenstein}\ \emph {et~al.}(2017)\citenamefont
  {Lewenstein}, \citenamefont {Sanpera},\ and\ \citenamefont
  {Ahufinger}}]{mac-book}%
  \BibitemOpen
  \bibfield  {author} {\bibinfo {author} {\bibfnamefont {M.}~\bibnamefont
  {Lewenstein}}, \bibinfo {author} {\bibfnamefont {A.}~\bibnamefont {Sanpera}},
  \ and\ \bibinfo {author} {\bibfnamefont {V.}~\bibnamefont {Ahufinger}},\
  }\href {https://www.xarg.org/ref/a/0198785801/} {\emph {\bibinfo {title}
  {Ultracold Atoms in Optical Lattices: Simulating quantum many-body
  systems}}}\ (\bibinfo  {publisher} {Oxford University Press},\ \bibinfo
  {year} {2017})\BibitemShut {NoStop}%
\bibitem [{\citenamefont {Chanda}\ \emph {et~al.}(2016)\citenamefont {Chanda},
  \citenamefont {Das}, \citenamefont {Sadhukhan}, \citenamefont {Pal},
  \citenamefont {Sen(De)},\ and\ \citenamefont {Sen}}]{atxy_pra}%
  \BibitemOpen
  \bibfield  {author} {\bibinfo {author} {\bibfnamefont {T.}~\bibnamefont
  {Chanda}}, \bibinfo {author} {\bibfnamefont {T.}~\bibnamefont {Das}},
  \bibinfo {author} {\bibfnamefont {D.}~\bibnamefont {Sadhukhan}}, \bibinfo
  {author} {\bibfnamefont {A.~K.}\ \bibnamefont {Pal}}, \bibinfo {author}
  {\bibfnamefont {A.}~\bibnamefont {Sen(De)}}, \ and\ \bibinfo {author}
  {\bibfnamefont {U.}~\bibnamefont {Sen}},\ }\href {\doibase
  10.1103/PhysRevA.94.042310} {\bibfield  {journal} {\bibinfo  {journal} {Phys.
  Rev. A}\ }\textbf {\bibinfo {volume} {94}},\ \bibinfo {pages} {042310}
  (\bibinfo {year} {2016})}\BibitemShut {NoStop}%
\bibitem [{\citenamefont {Roy}\ \emph {et~al.}(2019)\citenamefont {Roy},
  \citenamefont {Chanda}, \citenamefont {Das}, \citenamefont {Sadhukhan},
  \citenamefont {Sen(De)},\ and\ \citenamefont {Sen}}]{amader_dm}%
  \BibitemOpen
  \bibfield  {author} {\bibinfo {author} {\bibfnamefont {S.}~\bibnamefont
  {Roy}}, \bibinfo {author} {\bibfnamefont {T.}~\bibnamefont {Chanda}},
  \bibinfo {author} {\bibfnamefont {T.}~\bibnamefont {Das}}, \bibinfo {author}
  {\bibfnamefont {D.}~\bibnamefont {Sadhukhan}}, \bibinfo {author}
  {\bibfnamefont {A.}~\bibnamefont {Sen(De)}}, \ and\ \bibinfo {author}
  {\bibfnamefont {U.}~\bibnamefont {Sen}},\ }\href {\doibase
  10.1103/PhysRevB.99.064422} {\bibfield  {journal} {\bibinfo  {journal} {Phys.
  Rev. B}\ }\textbf {\bibinfo {volume} {99}},\ \bibinfo {pages} {064422}
  (\bibinfo {year} {2019})}\BibitemShut {NoStop}%
\bibitem [{\citenamefont {Peres}(2006)}]{Peres}%
  \BibitemOpen
  \bibfield  {author} {\bibinfo {author} {\bibfnamefont {A.}~\bibnamefont
  {Peres}},\ }\href
  {https://www.amazon.com/Quantum-Theory-Concepts-Fundamental-Theories-ebook/dp/B000W7Y9C6?SubscriptionId=AKIAIOBINVZYXZQZ2U3A&tag=chimbori05-20&linkCode=xm2&camp=2025&creative=165953&creativeASIN=B000W7Y9C6}
  {\emph {\bibinfo {title} {Quantum Theory: Concepts and Methods (Fundamental
  Theories of Physics Book 57)}}}\ (\bibinfo  {publisher} {Springer},\ \bibinfo
  {year} {2006})\BibitemShut {NoStop}%
\bibitem [{\citenamefont {Vidal}\ and\ \citenamefont
  {Werner}(2002)}]{negativity}%
  \BibitemOpen
  \bibfield  {author} {\bibinfo {author} {\bibfnamefont {G.}~\bibnamefont
  {Vidal}}\ and\ \bibinfo {author} {\bibfnamefont {R.~F.}\ \bibnamefont
  {Werner}},\ }\href {\doibase 10.1103/PhysRevA.65.032314} {\bibfield
  {journal} {\bibinfo  {journal} {Phys. Rev. A}\ }\textbf {\bibinfo {volume}
  {65}},\ \bibinfo {pages} {032314} (\bibinfo {year} {2002})}\BibitemShut
  {NoStop}%
\bibitem [{\citenamefont {Plenio}(2005)}]{ln}%
  \BibitemOpen
  \bibfield  {author} {\bibinfo {author} {\bibfnamefont {M.~B.}\ \bibnamefont
  {Plenio}},\ }\href {\doibase 10.1103/PhysRevLett.95.090503} {\bibfield
  {journal} {\bibinfo  {journal} {Phys. Rev. Lett.}\ }\textbf {\bibinfo
  {volume} {95}},\ \bibinfo {pages} {090503} (\bibinfo {year}
  {2005})}\BibitemShut {NoStop}%
\bibitem [{\citenamefont {Peres}(1996)}]{ap-sep}%
  \BibitemOpen
  \bibfield  {author} {\bibinfo {author} {\bibfnamefont {A.}~\bibnamefont
  {Peres}},\ }\href {\doibase 10.1103/PhysRevLett.77.1413} {\bibfield
  {journal} {\bibinfo  {journal} {Phys. Rev. Lett.}\ }\textbf {\bibinfo
  {volume} {77}},\ \bibinfo {pages} {1413} (\bibinfo {year}
  {1996})}\BibitemShut {NoStop}%
\bibitem [{\citenamefont {Horodecki}\ \emph {et~al.}(1996)\citenamefont
  {Horodecki}, \citenamefont {Horodecki},\ and\ \citenamefont
  {Horodecki}}]{necess-suff}%
  \BibitemOpen
  \bibfield  {author} {\bibinfo {author} {\bibfnamefont {M.}~\bibnamefont
  {Horodecki}}, \bibinfo {author} {\bibfnamefont {P.}~\bibnamefont
  {Horodecki}}, \ and\ \bibinfo {author} {\bibfnamefont {R.}~\bibnamefont
  {Horodecki}},\ }\href {\doibase 10.1016/s0375-9601(96)00706-2} {\bibfield
  {journal} {\bibinfo  {journal} {Phys. Lett. A}\ }\textbf {\bibinfo {volume}
  {223}},\ \bibinfo {pages} {1} (\bibinfo {year} {1996})}\BibitemShut {NoStop}%
\bibitem [{\citenamefont {Cover}\ and\ \citenamefont
  {Thomas}(2006)}]{CoverThomas}%
  \BibitemOpen
  \bibfield  {author} {\bibinfo {author} {\bibfnamefont {T.~M.}\ \bibnamefont
  {Cover}}\ and\ \bibinfo {author} {\bibfnamefont {J.~A.}\ \bibnamefont
  {Thomas}},\ }\href@noop {} {\emph {\bibinfo {title} {Elements of Information
  Theory (Wiley Series in Telecommunications and Signal Processing)}}}\
  (\bibinfo  {publisher} {Wiley-Interscience},\ \bibinfo {address} {New York,
  NY, USA},\ \bibinfo {year} {2006})\ pp.\ \bibinfo {pages}
  {19--25}\BibitemShut {NoStop}%
\bibitem [{\citenamefont {Henderson}\ and\ \citenamefont
  {Vedral}(2001)}]{totalcorr1}%
  \BibitemOpen
  \bibfield  {author} {\bibinfo {author} {\bibfnamefont {L.}~\bibnamefont
  {Henderson}}\ and\ \bibinfo {author} {\bibfnamefont {V.}~\bibnamefont
  {Vedral}},\ }\href {\doibase 10.1088/0305-4470/34/35/315} {\bibfield
  {journal} {\bibinfo  {journal} {J Phys A: Mathematical and General}\ }\textbf
  {\bibinfo {volume} {34}},\ \bibinfo {pages} {6899} (\bibinfo {year}
  {2001})}\BibitemShut {NoStop}%
\bibitem [{\citenamefont {Groisman}\ \emph {et~al.}(2005)\citenamefont
  {Groisman}, \citenamefont {Popescu},\ and\ \citenamefont
  {Winter}}]{totalcorr2}%
  \BibitemOpen
  \bibfield  {author} {\bibinfo {author} {\bibfnamefont {B.}~\bibnamefont
  {Groisman}}, \bibinfo {author} {\bibfnamefont {S.}~\bibnamefont {Popescu}}, \
  and\ \bibinfo {author} {\bibfnamefont {A.}~\bibnamefont {Winter}},\ }\href
  {\doibase 10.1103/PhysRevA.72.032317} {\bibfield  {journal} {\bibinfo
  {journal} {Phys. Rev. A}\ }\textbf {\bibinfo {volume} {72}},\ \bibinfo
  {pages} {032317} (\bibinfo {year} {2005})}\BibitemShut {NoStop}%
\bibitem [{\citenamefont {Lieb}\ \emph {et~al.}(1961)\citenamefont {Lieb},
  \citenamefont {Schultz},\ and\ \citenamefont {Mattis}}]{lsm}%
  \BibitemOpen
  \bibfield  {author} {\bibinfo {author} {\bibfnamefont {E.}~\bibnamefont
  {Lieb}}, \bibinfo {author} {\bibfnamefont {T.}~\bibnamefont {Schultz}}, \
  and\ \bibinfo {author} {\bibfnamefont {D.}~\bibnamefont {Mattis}},\ }\href
  {\doibase 10.1016/0003-4916(61)90115-4} {\bibfield  {journal} {\bibinfo
  {journal} {Ann. Phys.}\ }\textbf {\bibinfo {volume} {16}},\ \bibinfo {pages}
  {407} (\bibinfo {year} {1961})}\BibitemShut {NoStop}%
\bibitem [{\citenamefont {Barouch}\ and\ \citenamefont
  {McCoy}(1971{\natexlab{a}})}]{bm2}%
  \BibitemOpen
  \bibfield  {author} {\bibinfo {author} {\bibfnamefont {E.}~\bibnamefont
  {Barouch}}\ and\ \bibinfo {author} {\bibfnamefont {B.~M.}\ \bibnamefont
  {McCoy}},\ }\href {\doibase 10.1103/PhysRevA.3.786} {\bibfield  {journal}
  {\bibinfo  {journal} {Phys. Rev. A}\ }\textbf {\bibinfo {volume} {3}},\
  \bibinfo {pages} {786} (\bibinfo {year} {1971}{\natexlab{a}})}\BibitemShut
  {NoStop}%
\bibitem [{\citenamefont {Barouch}\ and\ \citenamefont
  {McCoy}(1971{\natexlab{b}})}]{bm3}%
  \BibitemOpen
  \bibfield  {author} {\bibinfo {author} {\bibfnamefont {E.}~\bibnamefont
  {Barouch}}\ and\ \bibinfo {author} {\bibfnamefont {B.~M.}\ \bibnamefont
  {McCoy}},\ }\href {\doibase 10.1103/PhysRevA.3.2137} {\bibfield  {journal}
  {\bibinfo  {journal} {Phys. Rev. A}\ }\textbf {\bibinfo {volume} {3}},\
  \bibinfo {pages} {2137} (\bibinfo {year} {1971}{\natexlab{b}})}\BibitemShut
  {NoStop}%
\bibitem [{\citenamefont {Yu}\ and\ \citenamefont {Eberly}(2007)}]{x1}%
  \BibitemOpen
  \bibfield  {author} {\bibinfo {author} {\bibfnamefont {T.}~\bibnamefont
  {Yu}}\ and\ \bibinfo {author} {\bibfnamefont {J.~H.}\ \bibnamefont
  {Eberly}},\ }\href {http://www.rintonpress.com/xqic7/qic-7-56/459-468.pdf}
  {\bibfield  {journal} {\bibinfo  {journal} {Quantum Information {\&}
  Computation}\ }\textbf {\bibinfo {volume} {7}},\ \bibinfo {pages} {459}
  (\bibinfo {year} {2007})}\BibitemShut {NoStop}%
\bibitem [{\citenamefont {Mendonça}\ \emph {et~al.}(2014)\citenamefont
  {Mendonça}, \citenamefont {Marchiolli},\ and\ \citenamefont {Galetti}}]{x2}%
  \BibitemOpen
  \bibfield  {author} {\bibinfo {author} {\bibfnamefont {P.~E.}\ \bibnamefont
  {Mendonça}}, \bibinfo {author} {\bibfnamefont {M.~A.}\ \bibnamefont
  {Marchiolli}}, \ and\ \bibinfo {author} {\bibfnamefont {D.}~\bibnamefont
  {Galetti}},\ }\href {\doibase https://doi.org/10.1016/j.aop.2014.08.017}
  {\bibfield  {journal} {\bibinfo  {journal} {Annals of Physics}\ }\textbf
  {\bibinfo {volume} {351}},\ \bibinfo {pages} {79 } (\bibinfo {year}
  {2014})}\BibitemShut {NoStop}%
\end{thebibliography}%

\end{document}